\documentclass[a4paper,fleqn,usenatbib]{mnras}

\usepackage{newtxtext,newtxmath}

\usepackage[T1]{fontenc}
\usepackage{ae,aecompl}

\usepackage{graphicx}	
\usepackage{amsmath}	
\usepackage{amssymb}	
\usepackage{enumitem}
\usepackage{lineno}
\usepackage{comment}

\newcommand{\lsim}{\protect\raisebox{-0.8ex}{$\:\stackrel{\textstyle <}{\sim}\:$}}

\newcommand{\hMpc}{\ensuremath{h^{-1}\ {\rm Mpc}}}
\newcommand{\hMpck}{\ensuremath{h\ {\rm Mpc}^{-1}}}
\newcommand{\hSM}{\ensuremath{h^{-1}\ M_{\odot}}}
\newcommand{\SM}{\ensuremath{M_{\odot}}}

\defcitealias{2018MNRAS.478.1866H}{HE18}

\title[Iterative reconstruction and galaxy fields]{Application of the iterative reconstruction to simulated galaxy fields}

\author[R. Hada et al.]{
Ryuichiro Hada$^{1, 2, 3}$\thanks{E-mail: ryuichiro.hada@cfa.harvard.edu}
and Daniel J. Eisenstein$^{1}$
\\
$^{1}$Harvard-Smithsonian Center for Astrophysics, 60 Garden St., Cambridge, MA 02138, USA\\
$^{2}$Astronomical Institute, Tohoku University, Aoba-ku, Sendai 980-8578, Japan\\
$^{3}$Division for Interdisciplinary Advanced Research and Education, Tohoku University, Aoba-ku, Sendai 980-8578, Japan
}

\date{Accepted XXX. Received YYY; in original form ZZZ}

\pubyear{\the\year}

\begin{document}
\label{firstpage}
\pagerange{\pageref{firstpage}--\pageref{lastpage}}
	\maketitle

\begin{abstract} 
We apply an iterative reconstruction method to galaxy mocks in redshift space obtained from $N$-body simulations. Comparing the two-point correlation functions for the reconstructed density field, we find that although the performance is limited by shot noise and galaxy bias compared to the matter field, the iterative method can still reconstruct the initial linear density field from the galaxy field better than the standard method both in real and in redshift space. Furthermore, the iterative method is able to reconstruct both the monopole and quadrupole more precisely, unlike the standard method. We see that as the number density of galaxies gets smaller, the performance of reconstruction gets worse due to the sparseness. However, the precision in the determination of bias ($\sim20\%$) hardly impacts on the reconstruction processes.
\end{abstract}


\section{Introduction} \label{sec:intro}

Extracting the cosmological information from biased observables (i.e. galaxies, quasars, etc.) tracing the non-linear matter density distribution is one of the most puzzling problems in studying the large-scale structure (LSS). The baryon acoustic oscillation (BAO) peak in the two-point correlation function of galaxies~\citep[][]{1970ApJ...162..815P,1970Ap&SS...7....3S,2005MNRAS.362..505C,2005ApJ...633..560E} is well known as a standard ruler to measure the distance to the galaxy  sample, which allows us to understand the nature of spacetime and constrain the cosmological parameters~ \citep[e.g.,][]{2013PhR...530...87W}. However, the nonlinearity, sparseness, and clustering bias in the observed galaxy distribution makes the distance measurement with the BAO more complicated.

As the Universe evolves, each galaxy is differently moved by large-scale flows and then the BAO feature is smeared out.  This reduces the accuracy of the BAO distance measurement~\citep[][]{1999MNRAS.304..851M,2005Natur.435..629S,2005MNRAS.362L..25A,2005ApJ...633..575S,2006ApJ...651..619J,2007APh....26..351H,2008MNRAS.383..755A}. Fortunately, we can estimate the displacement of galaxies from the observed galaxy distribution, i.e., gravitational potential field. Therefore, moving galaxies back to the initial position, we are able to recover the BAO peak~\citep{2007ApJ...664..660E,2007ApJ...664..675E}, which is called standard reconstruction. 

While the standard reconstruction technique remains a simple and strong technique for restoring the BAO peak~\citep[][]{2012MNRAS.427.2132P,2012MNRAS.427.3435A,2014MNRAS.441...24A,2016arXiv161003506V,2017MNRAS.464.1168R}, there are some problems to be improved: 1) lacking other information, the standard method uses the final galaxy density field instead of the initial linear density field to estimate the displacements of galaxies. 2) the standard method takes account of 1st order perturbations only~\citep[the Zel'dovich approximation,][]{1970A&A.....5...84Z} when estimating the displacement. 3) the Kaiser formula~\citep[][]{1987MNRAS.227....1K}, which is used to model the redshift-space distortions in the standard method, doesn't fully capture the redshift-space distortions of the Zel'dovich approximation. 4) in the standard reconstruction procedure, the galaxies and the random particles in redshift space are displaced by different displacements in order to partially enforce the Kaiser approximation to redshift-space distortions. This results in $O(1)$ fictitious density fluctuations wherever the survey selection function varies quickly~\citepalias[for the detail, see Section~2.2 in][]{2018MNRAS.478.1866H}.

In ongoing and future galaxy redshift surveys, such as DESI \citep[][]{2016arXiv161100036D}, PFS \citep[][]{2014PASJ...66R...1T}, and Euclid \citep[][]{2011arXiv1110.3193L}, the BAO distance measurement is expected to be significantly improved by making the advantage of the width and depth of the survey. Therefore we need to manage these problems that the standard reconstruction has for making the BAO measurement more precise and reliable. Taking it into account, in our previous paper \citep[][hereafter HE18]{2018MNRAS.478.1866H}, we developed a new iterative reconstruction method, motivated by \citet{1999MNRAS.308..763M}. This iterative method can successfully address some of the above problems, especially 1) and 4). Furthermore, applying it to the simulated matter density field, we found that our method can estimate the displacement and restore the two-point correlation function, both in real and in redshift space, more successfully compared with the standard method.  

In addition, there are some different types of iterative reconstruction methods~\citep[e.g.,][]{2010ApJ...720.1650S,2012JCAP...10..006T,2017PhRvD..96l3502Z,2017PhRvD..96b3505S,2018PhRvD..97b3505S}. The main qualitative difference between our method and other iterative methods is that we take account of only the 1st order in LPT (simpler) while making the solution converge in the iteration process (more reliable). We gave a more comprehensive comparison with these previous works in Section 3.4 of \citet[][]{2018MNRAS.478.1866H}.

In this paper, we apply the iterative method to simulated biased tracer. The reconstruction from biased tracers have been studied \citep[e.g.,][]{2009PhRvD..80l3501N,2017ApJ...847..110Y,2018arXiv180706381W,2018arXiv180908135B}, and it is found that although the reconstruction is still able to restore the BAO peak reasonably, the performance is largely limited by shot noise and clustering bias compared to the matter field. In addition, the impact of the sparseness and bias in biased tracers on the distance measurement has been investigated so far \citep[e.g.,][]{2011ApJ...734...94M,2014MNRAS.445.3152B}, which showed that the efficiency of reconstruction is increased as the number density of galaxies is increased (the bias gets close to 1 accordingly). In order to evaluate the performance of our iterative reconstruction for biased tracers, we will test it on the simulated galaxy fields, with various parameter settings (e.g., smoothing scales, number densities, etc.)   
The paper is organized as follows: we introduce the iterative reconstruction method that was introduced by \citetalias{2018MNRAS.478.1866H} in Section~\ref{sec:method}. 
Section~\ref{sec:simulations} summarizes the simulated galaxy samples. In Section~\ref{sec:Result}, we see the utility of anisotropic smoothing in redshift space, and then compare the correlation functions reconstructed from the galaxy fields, changing some parameters. Finally, we summarize our conclusion in Section~\ref{sec:Conc}.

\section{Reconstruction method} \label{sec:method}

\subsection{\label{sec:Iterative}Iterative reconstruction}

We begin by introducing the iterative reconstruction method that we proposed in \citetalias{2018MNRAS.478.1866H}, which was motivated by \citet{1999MNRAS.308..763M}.  

Lagrangian perturbation theory (LPT) describes the dynamics of objects in terms of the displacement field ${\bf S}$ from the initial position ${\bf q}$ to the final Eulerian position ${\bf x}$:
\begin{eqnarray}
	{\bf x}({\bf q}, t)	= {\bf q} + {\bf S}({\bf q}, t).  \label{eq:x_q}
\end{eqnarray}
Here we opt to use only first-order displacements, ${\bf S} = {\bf S}^{(1)}$, for simplicity and because we expect that the sparseness of realistic galaxy samples will require large enough smoothing scales that  first-order will be sufficient. Then, we can describe the redshift-space displacement ${\bf S}^{(s)}$ by
\begin{eqnarray}
	{\bf s}({\bf q}, t) = {\bf q} + {\bf S}^{(s)}({\bf q}, t),   \label{eq:s2}
\end{eqnarray}
where 
\begin{eqnarray}
	{\bf S}^{(s)}({\bf q}, t) = {\bf S}^{(1)} + f[{\bf S}^{(1)} \cdot \hat{\bf z}]\hat{\bf z}.  \label{eq:S^s}
\end{eqnarray}
Here $f = {\rm d} \ln D/ {\rm d} \ln a$ is the linear growth rate where $D$ is the linear growth factor. 

Here we assume that the linear density contrast $\delta_{\rm L}$ can be separated into the large-scale part $\delta_{l}$ and residual part $\delta_{\rm res}$:
\begin{eqnarray}
	\delta_{\rm L}({\bf q},t) = \delta_{l}({\bf q},t) + \delta_{\rm res}({\bf q}).
	\label{eq:separation}
\end{eqnarray}
We consider a model in which only the large-scale portion creates displacements, which then advect the small-scale residual as a passive tracer. 
While this is obviously not correct on small scales, by using a smoothing filter to do the scale separation, we create a smooth transition between the two regimes. In particular, we define the large scale field via
\begin{eqnarray}
	\tilde{\bf S}^{(1)}_{l}({\bf k}) &=& \frac{i {\bf k}}{k^2} \tilde{\delta}_{\rm L}({\bf k}) G(k),  \label{eq:shift_large} \\
	\delta_{l}({\bf q},t) &=& - \nabla \cdot {\bf S}_{l}^{(1)}({\bf q},t).
	\label{eq:large}
\end{eqnarray}
Here the tilde over scalars and vectors denotes that the quantities are in Fourier space. Eq.~(\ref{eq:shift_large}) corresponds to the linear solution in LPT (the Zel'dovich approximation). The assumption above means that the residual part is assumed to have existed at the initial time: $\rho({\bf q}) = \bar{\rho}(1 +  \delta_{\rm res}({\bf q}))$. The continuity equation is then described as follows:
\begin{eqnarray}
	\det \left[\delta^{\rm K}_{ab} + S^{(s)}_{l|a, b} \right] = \frac{\rho({\bf q})}{\rho({\bf s})} = \frac{1 + \delta_{\rm res}({\bf q})}{1 + \delta_{s}({\bf s})}.  \label{eq:continuity_mo}
\end{eqnarray}
where $\delta^{\rm K}_{ab}$ is the Kronecker delta, $\delta_{s}({\bf s})$ is the observed density in redshift space, and ${\bf S}_{l}^{(s)}$ is related to ${\bf S}_{l}^{(1)}$ through Eq.~(\ref{eq:S^s}).

Furthermore, to mitigate the effect of redshift-space distortions coming from small-scale thermal motions, i.e., the Fingers of God effect, we seek to down-weight these density fluctuations in the smoothed density field. We therefore introduce the parameter, $\mathcal{C}_{\rm ani}$, as follows: 
\begin{eqnarray}
	\mathcal{C}_{\rm ani} &\equiv& \Sigma_{\parallel}/\Sigma_{\perp}, \label{eq:C_ani}
\end{eqnarray}
where $\Sigma_{\parallel}$ and $\Sigma_{\perp}$ are the smoothing scales along the line of sight and the perpendicular directions, respectively:
\begin{eqnarray}
	G_{\rm ani}(k) = \exp[- 0.5 (k_{\perp}^{2} + k_{\parallel}^{2}\mathcal{C}_{\rm ani}^{2})\Sigma^{2}_{\perp}]. \label{eq:G_ani}
\end{eqnarray}
Effectively, this means that we use less of the line-of-sight density fluctuations in deriving the large-scale displacements.

Our final goal is to find the linear density field $\delta_{\rm L}({\bf q})$ that solves Eq.~(\ref{eq:continuity_mo}) given the observed density $\delta_{s}({\bf s})$ and the definitions in Eq.~(\ref{eq:S^s})-(\ref{eq:large}) (solving problem 1). To do so, we begin with assigning the galaxy particles on a grid and calculate the observed density field in redshift space $\delta_{s}({\bf s})$ at each grid cell. We then repeat steps estimating the displacement and updating the guess for the linear density at each grid cell (solving problem 4) until they converge. Further, as we are starting from the Lagrangian evolution of an initial field, the results seamlessly include the large-scale redshift-space distortions (problem 3) \citepalias[for the detail of our implementation, see Section~3.2 in][]{2018MNRAS.478.1866H}.

\subsection{\label{sec:app_gal}Application to galaxy fields}

Up to 1st order (in Eularian perturbation theory), the galaxy density field in redshift space $\tilde{\delta}^{g}_{s}$ (in Fourier space) is related to that in real space $\tilde{\delta}^{g}$~\citep{1987MNRAS.227....1K, 2002PhR...367....1B}: 
\begin{eqnarray}
	\tilde{\delta}_{s}^{g}({\bf k}) = (1 + \beta \mu^{2})\tilde{\delta}^{g}({\bf k}).  \label{eq:kaiser}
\end{eqnarray}
where $\mu = k_{z} / k$ ($z$ is the line-of-sight direction) and $\beta = f/b$. Here $b$ is the linear galaxy bias that is the ratio of the galaxy density to the matter density field: $\delta^{g} = b \delta$. Taking account of the fact that matter density fields correspond to $b = 1$, we need to replace the redshift-space density field and the linear growth rate as follows in the procedure, in applying our iterative method to galaxy fields: 
\begin{eqnarray}
	\delta_{s}({\bf s}) &\to& \delta^{g}_{s}({\bf s})/b,  \label{eq:dens_b}
	\\
	f &\to& \beta.  \label{eq:f_b}
\end{eqnarray}
As we mentioned in problem 3 of Section~\ref{sec:intro}, while the Kaiser formula is exactly correct as long as we consider only 1st order Eularian perturbation theory, we focus on the Lagrangian evolution of an initial field. Therefore, we should pay attention to how well the replacement above works for biased tracers.

\section{Simulations} \label{sec:simulations}

\subsection{\label{sec:matter}Matter density fields}

In this paper, we use $N$-body simulation data products from the {\sc Abacus} project~\citep[][]{2016MNRAS.461.4125G,2017arXiv171205768G}\footnote{\url{https://lgarrison.github.io/AbacusCosmos/}} to evaluate the performance of our iterative reconstruction method. {\sc Abacus} is an extremely fast and accurate $N$-body code for cosmological simulations and can compute over 100 billion pairwise force interactions per second on a single computer node~\citep{2018arXiv181002916G}. We use 15 emulator boxes with independent phases ({\tt emulator\_1100box\_planck\_00-\{1..15\}}) assuming \citet{2016A&A...594A..13P} as the fiducial cosmology, in which box size $L = 1100 \hMpc$, number of particles $N_{\rm p} = 1440^{3}$, and particle mass $M_{\rm p} \sim 4\times10^{10} \hSM$. 

Hereafter, we fix redshift for the observed non-linear density field to $z = 0.5$ and use about $480^{3}$ particles ($\sim 4\%$) chosen randomly from each realization as matter density fields.

\subsection{\label{sec:galaxy}Galaxy density fields}

To compare the performances between the matter and galaxy density fields, we need to create the galaxy catalog corresponding to the matter density fields defined in Section~\ref{sec:matter}. Each realization from emulator boxes includes the halo catalog characterized using {\sc Rockstar}~\citep[][]{2013ApJ...762..109B} halo finder. We then create the galaxy catalog from the halo catalogs with GeneRalized ANd Differentiable Halo Occupation Distribution~\citep[GRAND-HOD, ][]{2018MNRAS.tmp.1043Y},\footnote{\url{https://github.com/SandyYuan/GRAND-HOD}} which generalizes the standard 5 parameter halo occupation distribution model~\citep[HOD, ][]{2009ApJ...707..554Z, 2015ApJ...810...35K} with various halo-scale physics and assembly bias. In the standard HOD model, the average number of central and satellite galaxies, in a halo of mass $M$, is given by
\begin{eqnarray}
	\langle n_{\rm cen} \rangle & = & \frac{1}{2} \ {\rm erfc}\ \left[ \frac{\ln (M_{\rm cut}/M)}{\sqrt{2} \sigma} \right],  \label{eq:n_cen}
	\\
	\langle n_{\rm sat} \rangle & = & \left( \frac{M - \kappa M_{\rm cut}}{M_{1}} \right)^{\alpha},  \label{eq:n_sat}
\end{eqnarray}
where $M_{\rm cut}$ is the cut-off mass for the halo to host a central galaxy, $\sigma$ is the scatter around the cut-off mass, $\kappa M_{\rm cut}$ is the cut-off mass for the halo to host a satellite galaxy, $M_{1}$ is the typical mass scale for a halo to host one satellite, and $\alpha$ is the slope of the power-law for the number of satellites at high mass. 

When running GRAND-HOD, the values of 5 parameters above are set to $M_{\rm cut} = 10^{13.35} \SM$, $M_{1} = 10^{13.8} \SM$, $\sigma = 0.85$, $\alpha = 1$, and $\kappa = 1$, which are fitted to the two-point auto-correlation functions for Luminous Red Galaxies (LRGs) in the SDSS~\citep[][]{2005ApJ...621...22Z} and the two-point cross-correlation functions between the SDSS LRGs and galaxies in the SDSS imaging sample~\citep[][]{2005ApJ...619..178E}. In this setting, the number of galaxies is $\sim 84^3$ for each realization and the number density of galaxies $n_{\rm gal}$ is $\sim 4\times10^{-4}(\hMpc)^{-3}$. Note that we don't manipulate generalization parameters beyond the standard 5 parameter (all of them are set to $0$) because for now, we are interested in the performance of the iterative reconstruction for the simple and typical galaxy catalog.

\begin{figure*}
\begin{tabular}{ccc}
 \begin{minipage}{0.33\hsize}
  \begin{center}
   \includegraphics[width=65mm]{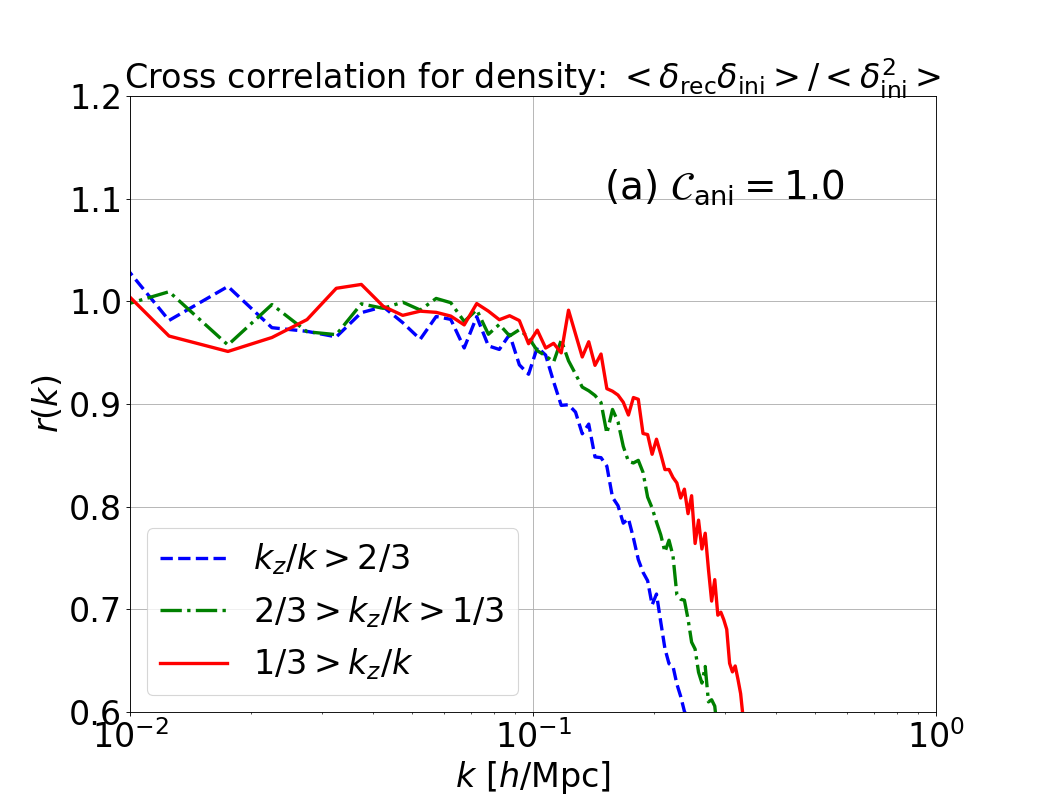}
  \end{center}
\end{minipage} 
 \begin{minipage}{0.33\hsize}
  \begin{center}
   \includegraphics[width=65mm]{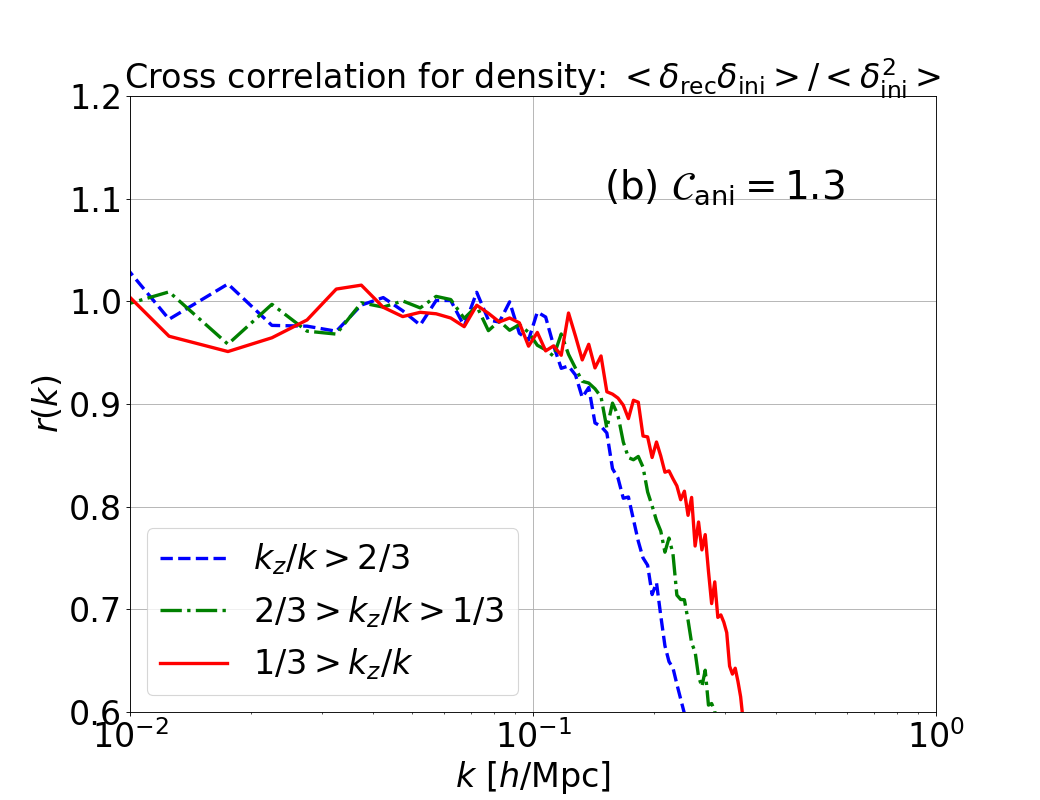}
  \end{center}
\end{minipage}
 \begin{minipage}{0.33\hsize}
  \begin{center}
   \includegraphics[width=65mm]{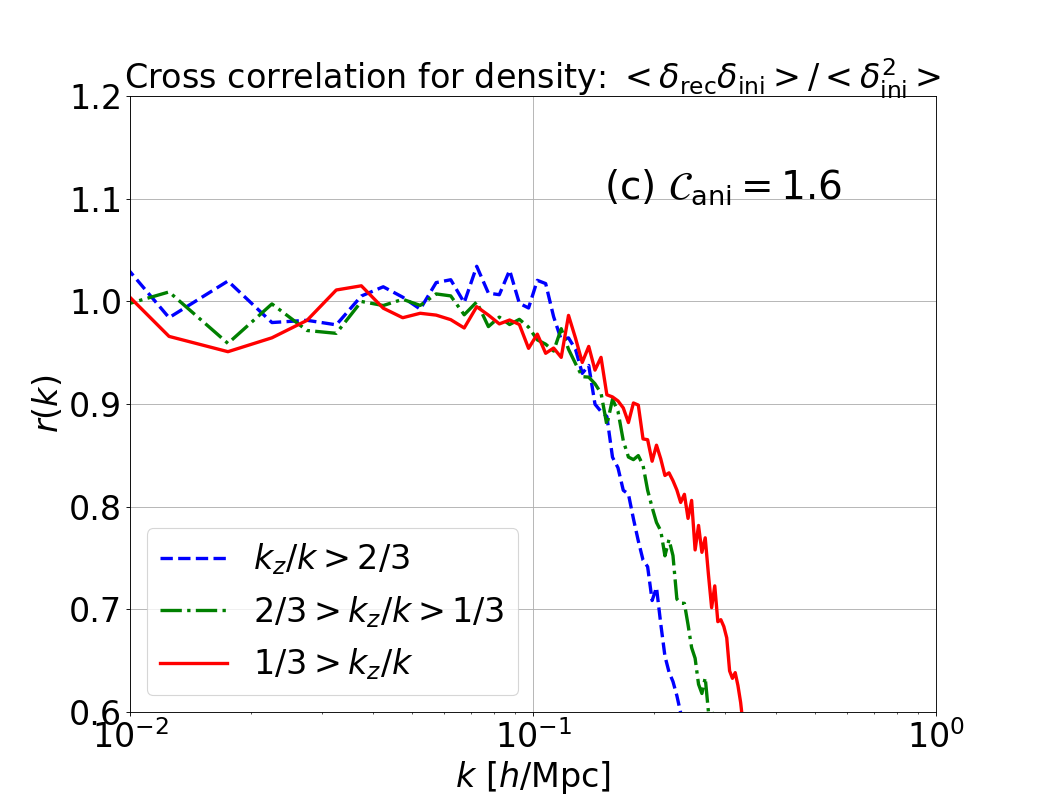}
  \end{center}
\end{minipage}
\end{tabular}
  \caption{\label{fig:sm_ani} Cross-correlation coefficient with the initial density field, $s[\delta^{g}_{\rm L}](k)$, in redshift space for various $\mathcal{C}_{\rm ani}$: 1.0 (a), 1.3 (b), and 1.6 (c). The solid (in red), dot-dashed (in green), and dashed (in blue) lines show the coefficients for $\Sigma_{\rm eff}  = 10\hMpc$ at $z = 0.5$ averaged over the directions: $1/3 > k_{z} / k$, $2/3 > k_{z} / k > 1/3$, and $k_{z} /k > 2/3$, respectively. In the case with $\mathcal{C}_{\rm ani} = 1.3$, all three directions are consistent with each other on large scales $k \lsim 0.1\hMpck$}
\end{figure*}

\section{Results} \label{sec:Result}

\subsection{\label{sec:params}Parameter setting}

Following discussions in Section~3.2 of \citetalias[][]{2018MNRAS.478.1866H}, we introduce some techniques to effectively make the solution converge and to avoid oscillations of the solution.

For the smoothing scale $\Sigma$, we start with a large initial value $\Sigma_{\rm ini}$ and reduce it gradually in the iterative process until it reaches the effective smoothing scale, $\Sigma_{\rm eff}$, that corresponds to the actual smoothing scale applied for the final displacement:
\begin{eqnarray}
	\Sigma_{\perp,n}  = \mbox{max}\left(\frac{\Sigma_{\rm ini}}{\mathcal{D}^{n}},\  \Sigma_{\rm eff}\right), \label{red_sm}
\end{eqnarray}
where $\Sigma_{\perp,n}$ is the $n$th smoothing scales along the perpendicular direction. We assume a constant ($>1$) as $\mathcal{D}$.  

Furthermore, in order to suppress oscillations in the iteration procedure, we weight the current value and the previous value:
\begin{eqnarray}
	\delta_{\rm L}^{(n)} = w \delta_{\rm L[ori]}^{(n)} + (1-w)\delta_{\rm L}^{(n-1)},  \label{eq:weight}
\end{eqnarray}
where $\delta_{\rm L}^{(n)}$ is the $n$th guess of the linear density, $\delta_{\rm L[ori]}^{(n)}$ is the (current) original guess, and $w$ is the weight: $0 < w <1$. 

In addition, we check the convergence of the solution by the ratio of the change in the guesses of the linear density to the observed density:
\begin{eqnarray}
	r_{\rm con} \equiv \frac{\sum [\delta_{\rm L[ori]}^{(n)} - \delta_{\rm L}^{(n-1)}]^{2} }{\sum \delta_{s}^{2}},  \label{eq:con}
\end{eqnarray}
where $\sum$ is the summation over all grid cells. Hereafter, we adopt $r_{\rm con} < 0.01$ as a criteria for convergence. 

In the following, we perform the reconstruction of the density field and the calculation of some types of correlations using a $480^{3}$ grid, and the annealing parameters are fixed to $\Sigma_{\rm ini} = 20\hMpc$ and $\mathcal{D} = 1.2$. Note that these two parameters have no physical meaning and the converged results should be independent of small changes in the exact annealing steps. In addition, we try some types of the effective smoothing scales: 5, 7, 10, and 15$\hMpc$ and then need to set the weight  and the numbers of iteration $n_{\rm iter}$ for each case so that the convergence criterion is satisfied. We summarize, in Table~\ref{tab:params}, the weight and the number of iteration for each effective smoothing scale that we used in the process of the iterative reconstruction. We emphasize that these two parameters are likely application-specific. 

Though in practice, we lack the information about the linear growth rate $f$ (or $\beta$) before reconstruction, we are interested in how well the iterative method can manage the effect of redshift-space distortions. Then, using a good approximation given by \citet{1991MNRAS.251..128L}, we adopt, as a fiducial $f$, the value for the fiducial cosmological parameters~\citep{2016A&A...594A..13P} that were assumed in the $N$-body simulations.

\begin{table}
  \begin{center}
	\caption{\label{tab:params} Weight and number of iteration for each smoothing scale}
    \begin{tabular}{ccc} \hline
      $\quad \Sigma_{\rm eff} \ (\hMpc) \quad$ & $\quad w \quad $& $\quad n_{\rm iter} \quad$
      \\ \hline 
          5     & 0.3 &  17         \\
          7     & 0.4 &  13         \\
        10     & 0.5 &    9         \\
        15     & 0.7 &    6                        
      \\ \hline
    \end{tabular}
  \end{center}
\end{table}

\subsection{\label{sec:x_corr}Cross correlation for density fields}

We define the cross-correlation coefficient in Fourier space between a density filed $\delta_{\rm X}$ and the initial density field $\delta_{\rm ini}$ as
\begin{eqnarray}
	s[\delta_{\rm X}](k) \equiv \frac{\langle \tilde{\delta}_{\rm X} \cdot  \tilde{\delta}^{*}_{\rm ini} \rangle}{ \langle |\tilde{\delta}_{\rm ini} |^2 \rangle}. \label{eq:coe_s}
\end{eqnarray}
Here $\delta_{\rm ini}$ is multiplied by the linear growth factor so as to extrapolate to the redshift corresponding to $\delta_{\rm X}$.

\subsubsection{\label{sec:bias_es}Bias estimation}

To apply our reconstruction method to galaxy (biased) fields, we need the galaxy bias $b$ in advance (see Eqs.~(\ref{eq:dens_b}) and (\ref{eq:f_b})). We then estimate the galaxy bias from the ratio of the cross-correlation coefficient between a galaxy field $\delta^{g}$ and the corresponding matter field $\delta$ (both the fields are measured in real space):
\begin{eqnarray}
	b = \mathcal{R}(k_{b}), \label{eq:b_R}
\end{eqnarray}
where
\begin{eqnarray}
	 \mathcal{R}(k) \equiv \frac{s[\delta^g](k)}{s[\delta](k)}. \label{eq:R_s}
\end{eqnarray}
Here $k_{b}$ is the reference wave number at which the value of the galaxy bias is decided. In this paper, we set the reference wave number $k_{b}$ to $0.1\hMpck$. Note that the value of bias we use hereafter is $\sim 2.3$ (though the exact values for each realization are slightly different from each other).

\begin{figure*}
\begin{tabular}{cc}
 \begin{minipage}{0.5\hsize}
  \begin{center}
   \includegraphics[width=80mm]{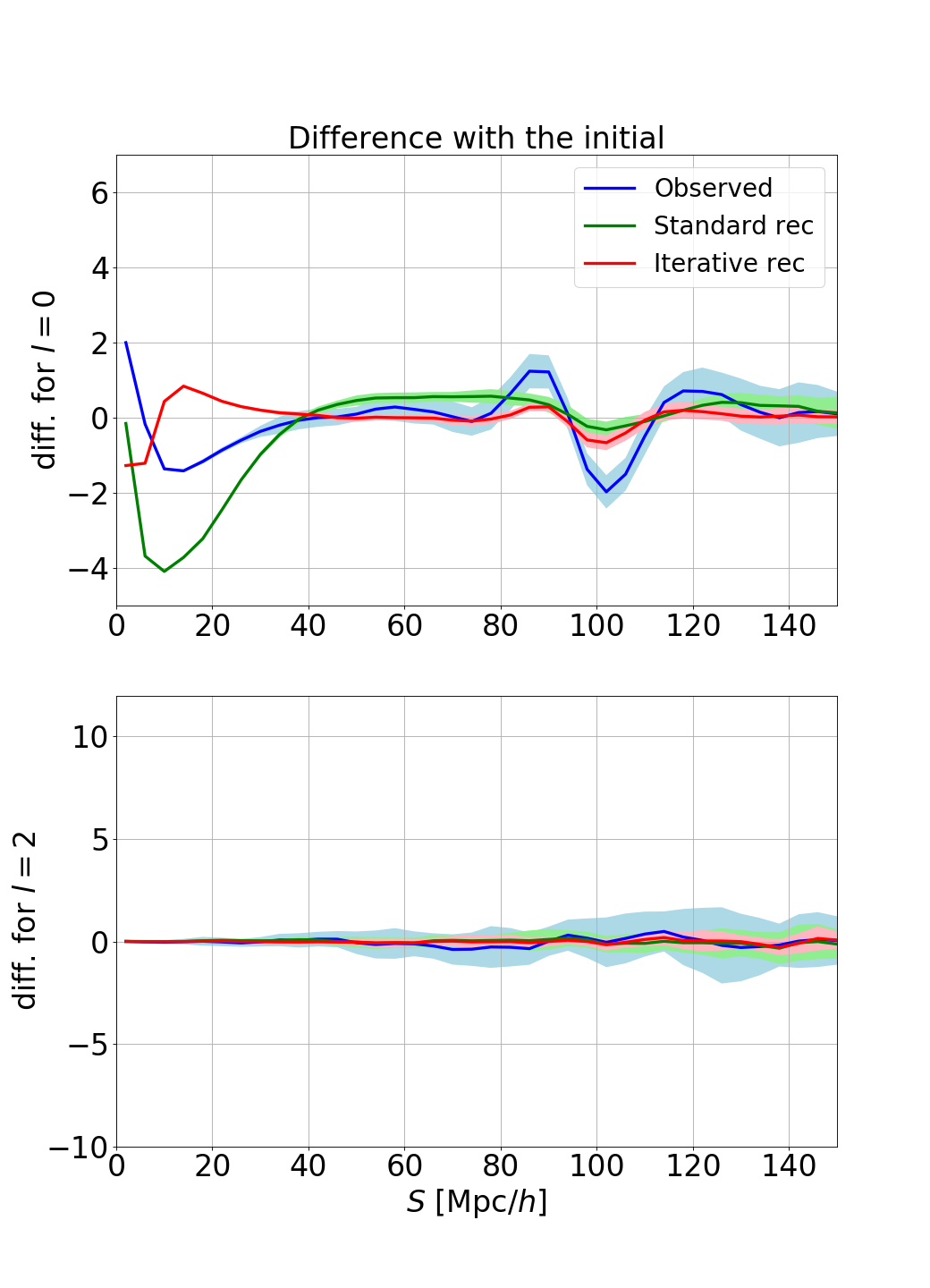}
  \end{center}
\end{minipage} 
 \begin{minipage}{0.5\hsize}
  \begin{center}
   \includegraphics[width=80mm]{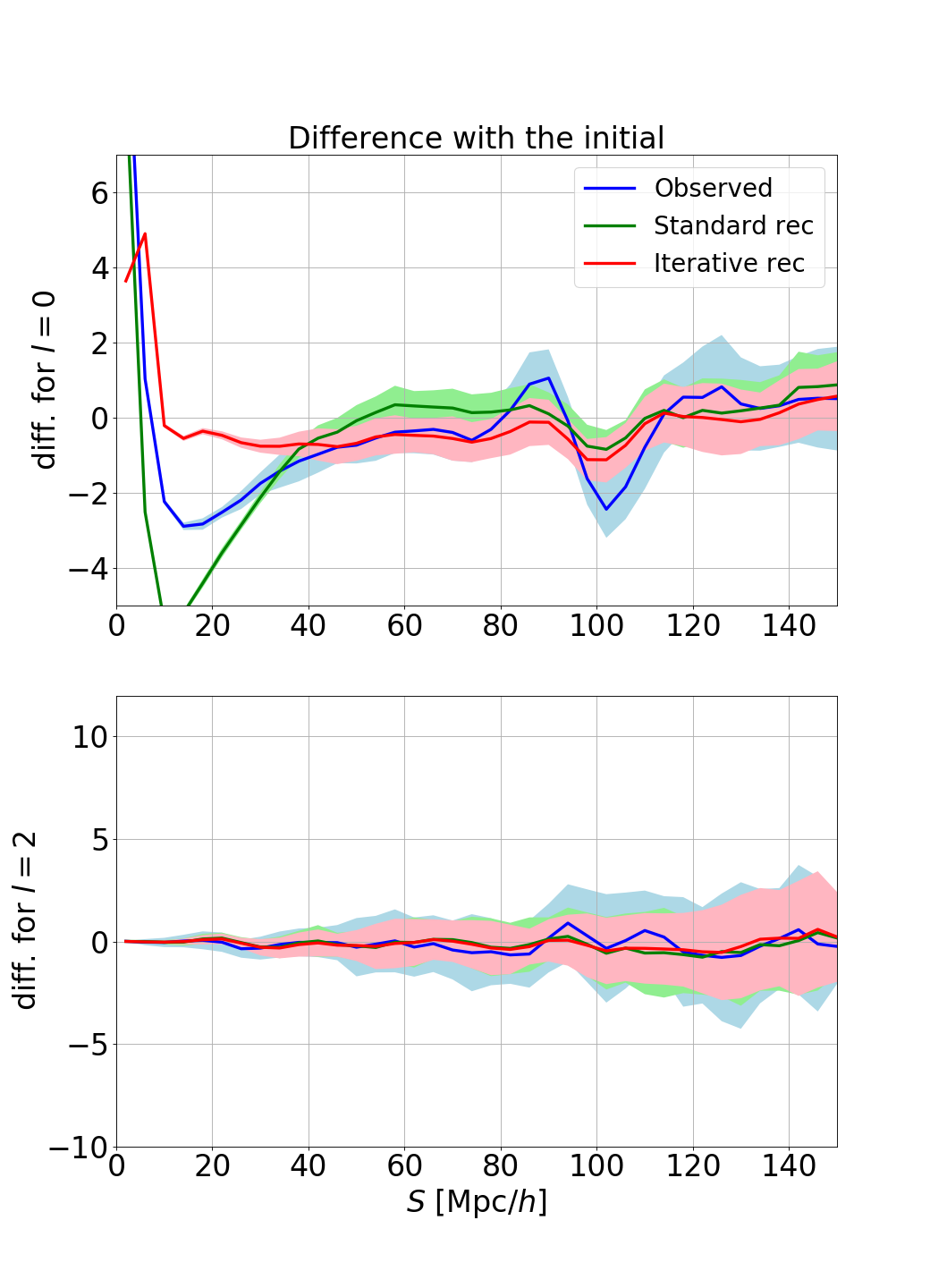}
  \end{center}
\end{minipage} \\
\begin{minipage}{0.5\hsize}
  \begin{center}
   \includegraphics[width=80mm]{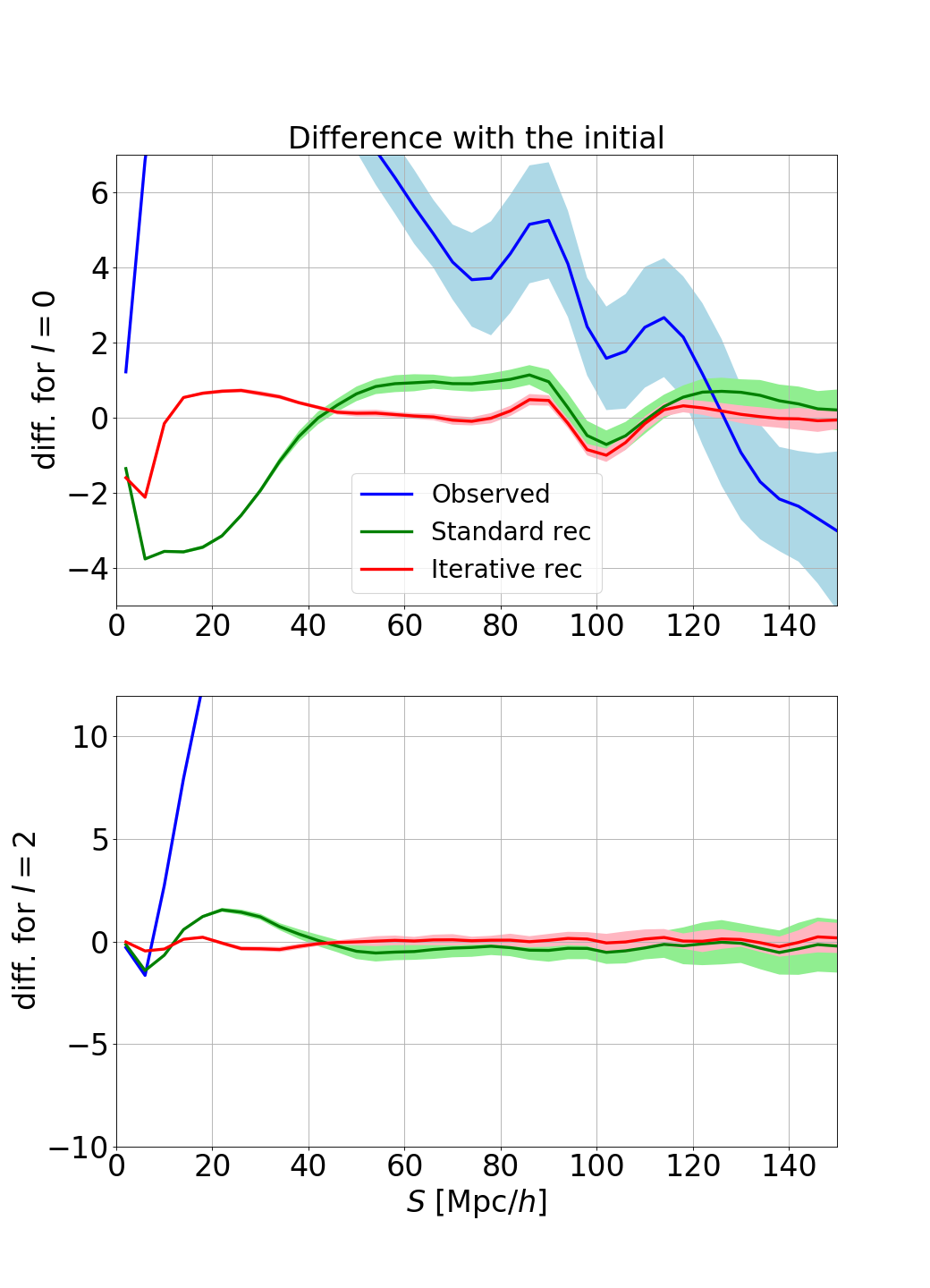}
  \end{center}
\end{minipage} 
 \begin{minipage}{0.5\hsize}
  \begin{center}
   \includegraphics[width=80mm]{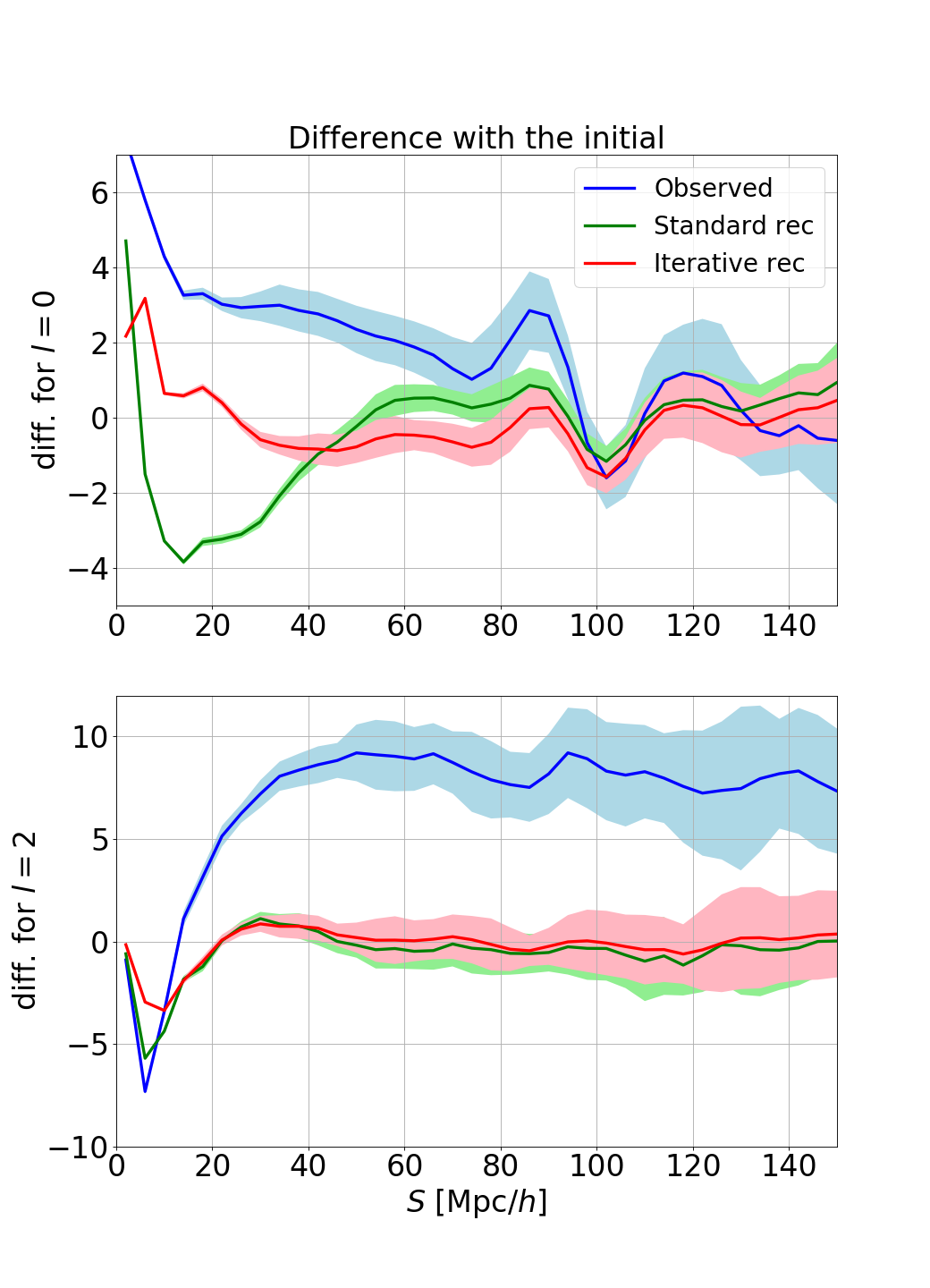}
  \end{center}
\end{minipage}
\end{tabular}
  \caption{\label{fig:MvsG}Comparison between the matter field and the galaxy field. {\it Left column}: matter field and {\it Right column}: galaxy field, in real space (upper row) and redshift space (lower row). The top panel and bottom panel in each row demonstrate the monopole ($l=0$) and the quadrupole ($l=2$) of $\Delta \xi_{l}[\delta_{\rm X}](S)$ (see Eq.~(\ref{eq:D_xi})), respectively. In each panel, the blue, green and red colors correspond to the observed density field (divided by the bias, $b$), the density field reconstructed with the standard method, and the one with our iterative method, respectively. While the solid line shows the average over 15 realizations, the shaded region displays the variance. Even for the galaxy density fields, our iterative method is still able to reconstruct the initial density field better than the standard method.}
\end{figure*}

\subsubsection{\label{sec:ani_sm}Anisotropic smoothing}

In Section~4.3 of \citetalias{2018MNRAS.478.1866H}, we discussed about the effect of redshift-space distortions coming from small-scale thermal motions, particularly the Finger of God effect and determined that for the matter density field in redshift space, the optimal value of parameter $\mathcal{C}_{\rm ani}$ is about 1.6. To estimate the optimal value of $\mathcal{C}_{\rm ani}$ for the galaxy 
density fields, we focus on the cross-correlation coefficient $s[\delta^{g}_{\rm L}]$, where $\delta^{g}_{\rm L}$ is the linear density contrast at $z = 0.5$ that is reconstructed from the galaxy field by using our iterative method with $\Sigma_{\rm eff} = 10\hMpc$. Note that the coefficient $s[\delta^{g}_{\rm L}]$ corresponds to $s(k)$, Eq.~(27), in \citetalias{2018MNRAS.478.1866H}.

Fig.~\ref{fig:sm_ani} shows $s[\delta^{g}_{\rm L}](k)$ with $\mathcal{C}_{\rm ani} = 1.0$, $1.3$ and $1.6$. Fig.~\ref{fig:sm_ani}a ($\mathcal{C}_{\rm ani} = 1.0$) corresponds to the reconstruction with an isotropic smoothing filter and the other two panels ($\mathcal{C}_{\rm ani} > 1.0$) show the results using less of the line-of-sight density fluctuations (see Eq.~\ref{eq:G_ani}). Compared with these results, we can see that the coefficient for the line of sight gets larger parallel to $\mathcal{C}_{\rm ani}$ and find that in Fig.~\ref{fig:sm_ani}b ($\mathcal{C}_{\rm ani} = 1.3$), the correlation coefficients for all wavevector directions are consistent with each other on large scales $k \lsim 0.1\hMpck$. 

Recalling that the optimal value of $\mathcal{C}_{\rm ani}$ for the matter density fields is 1.6, we find that in redshift space the matter fields are more influenced by the Finger of God effect than the galaxy fields. This result makes sense because the matter fields reflect the dynamics on smaller scales (inside halos) and we used a smaller smoothing scale, $\Sigma_{\rm eff} = 5\hMpc$, in applying our reconstruction method to the matter fields. In the following analysis in redshift space, we use the value $\mathcal{C}_{\rm ani} = 1.6$ for the matter fields and 1.3 for the galaxy fields.

\subsection{\label{sec:}Two-point correlation function}

In this section, we show the two-point correlation function to evaluate how well our iterative method works. The multiple moment of the two-point correlation is described as
\begin{eqnarray} 
\xi(S,\mu) = \sum^{\infty}_{l=0}\xi_{l}(S) P_{l}(\mu),   \label{eq:xi}
\end{eqnarray}
where $\xi(S,\mu)$ is the two-point correlation function for the density contrast and $P_{l}$ is the Legendre polynomial of order $l$. We are interested in the extent to which reconstruction methods can restore the initial linear density field and then define the difference between the two-point correlation function $\xi_{l}(S)$ for a density field $\delta_{\rm X}$ and the one for the initial density field in real space (multiplied by the linear growth factor) $\delta_{\rm ini}$ as 
\begin{eqnarray} 
	\Delta \xi_{l}[\delta_{\rm X}](S) \equiv \xi_{l}[\delta_{\rm X}](S) - \xi_{l}[\delta_{\rm ini}](S). \label{eq:D_xi}
\end{eqnarray}
Note that $\xi_{l}[\delta_{\rm ini}]$ here refers to an actual noisy realization in the simulations, not the ensemble averaged one over realizations. This means that $\Delta \xi_{l}[\delta_{\rm X}]$ above is defied for each realization of simulations.

\subsubsection{\label{sec:MvsG}Matter vs Galaxy}

The left column of Fig.~\ref{fig:MvsG} shows $\Delta \xi_{l}[\delta_{\rm L}](S)$ in real space (upper row) and redshift space (lower row), where $\delta_{\rm L}$ is the linear density contrast at $z = 0.5$ that is reconstructed from the matter field by using our iterative method (in red) and the standard method (in green) with $\Sigma_{\rm eff} = 5\hMpc$. The solid lines and the shaded regions show the average and the variance over 15 realizations, respectively. Compared with the results for the observed density field (in blue), we can see that in both real and redshift space, our iterative method successfully restore the monopole (top panel) and quadrupole (bottom panel) for the initial density field and make the dispersions over realizations (shaded region) smaller on all scales. Furthermore, focusing on the reconstructed monopole with both methods, we find that our iterative method considerably improves the standard reconstruction on scales $S \lsim 80 \hMpc$. In addition, we see that the quadrupole in redshift space is also restored more precisely by our method, especially on scales $S \lsim 40 \hMpc$ although there is no difference between the both methods in real space. 
 
The right column of Fig.~\ref{fig:MvsG} corresponds to $\Delta \xi_{l}[\delta^{g}_{\rm L}](S)$ in real space (upper row) and redshift space (lower row), where $\delta^{g}_{\rm L}$ is the linear density contrast at $z = 0.5$ that is reconstructed from the galaxy field with $\Sigma_{\rm eff} = 10\hMpc$. In both real and redshift space, we can see that our iterative method is still able to correctly reconstruct the initial density field. In addition, it is better than the standard method in the monopole, although there is no difference between both reconstruction methods in the quadrupole. We also find that the monopole reconstructed using our iterative method is systematically lower than the vertical line, $\Delta \xi_{l}(S) = 0$, which is caused by the uncertainty of the bias value (see also section~\ref{sec:bias}). Further, the galaxy field is much more noisy than the matter field because of the sparseness, exhibiting larger variance.

\begin{figure*}
\begin{tabular}{cc}
 \begin{minipage}{0.50\hsize}
  \begin{center}
   \includegraphics[width=80mm]{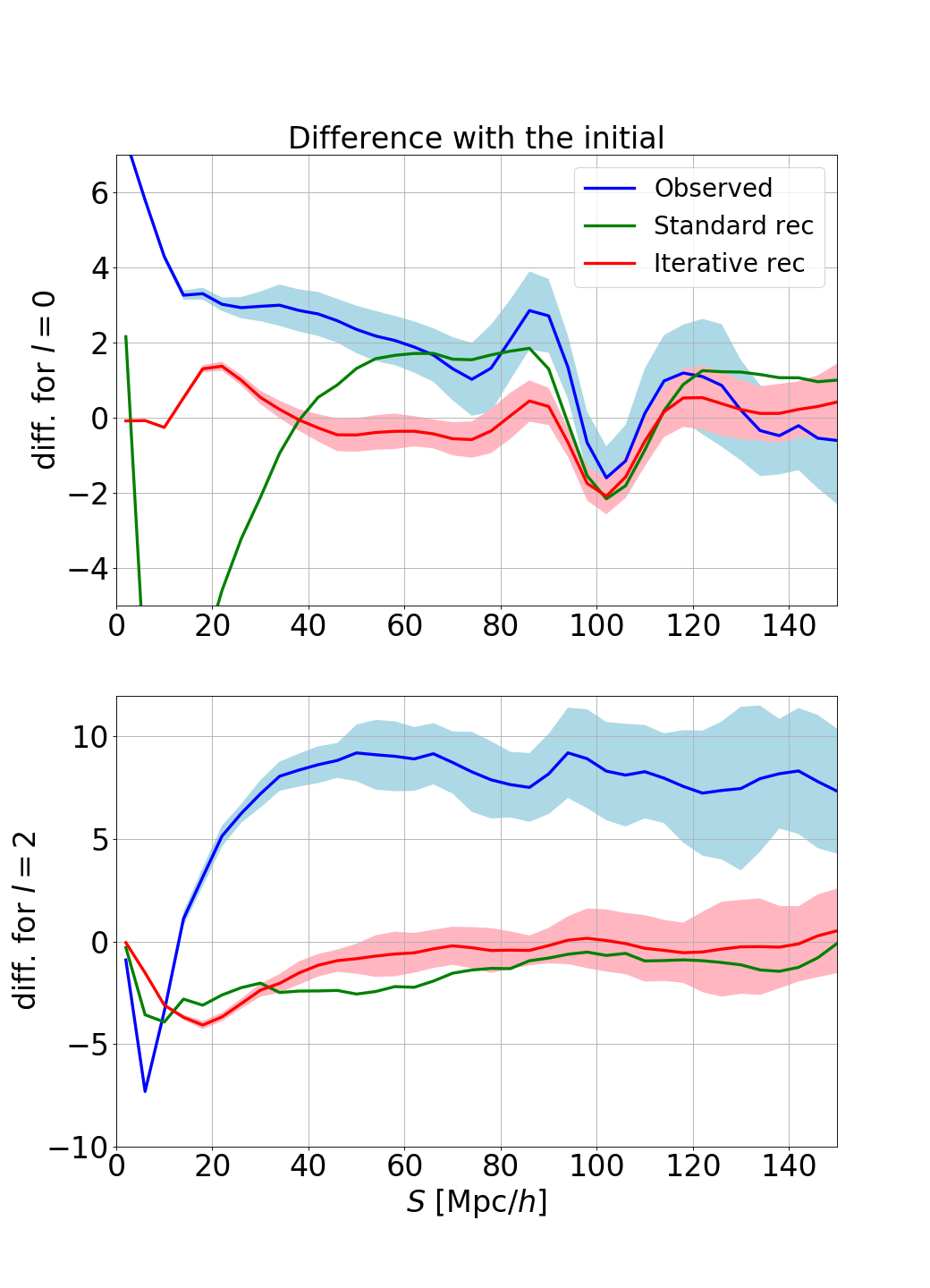}
  \end{center}
\end{minipage} 
 \begin{minipage}{0.50\hsize}
  \begin{center}
   \includegraphics[width=80mm]{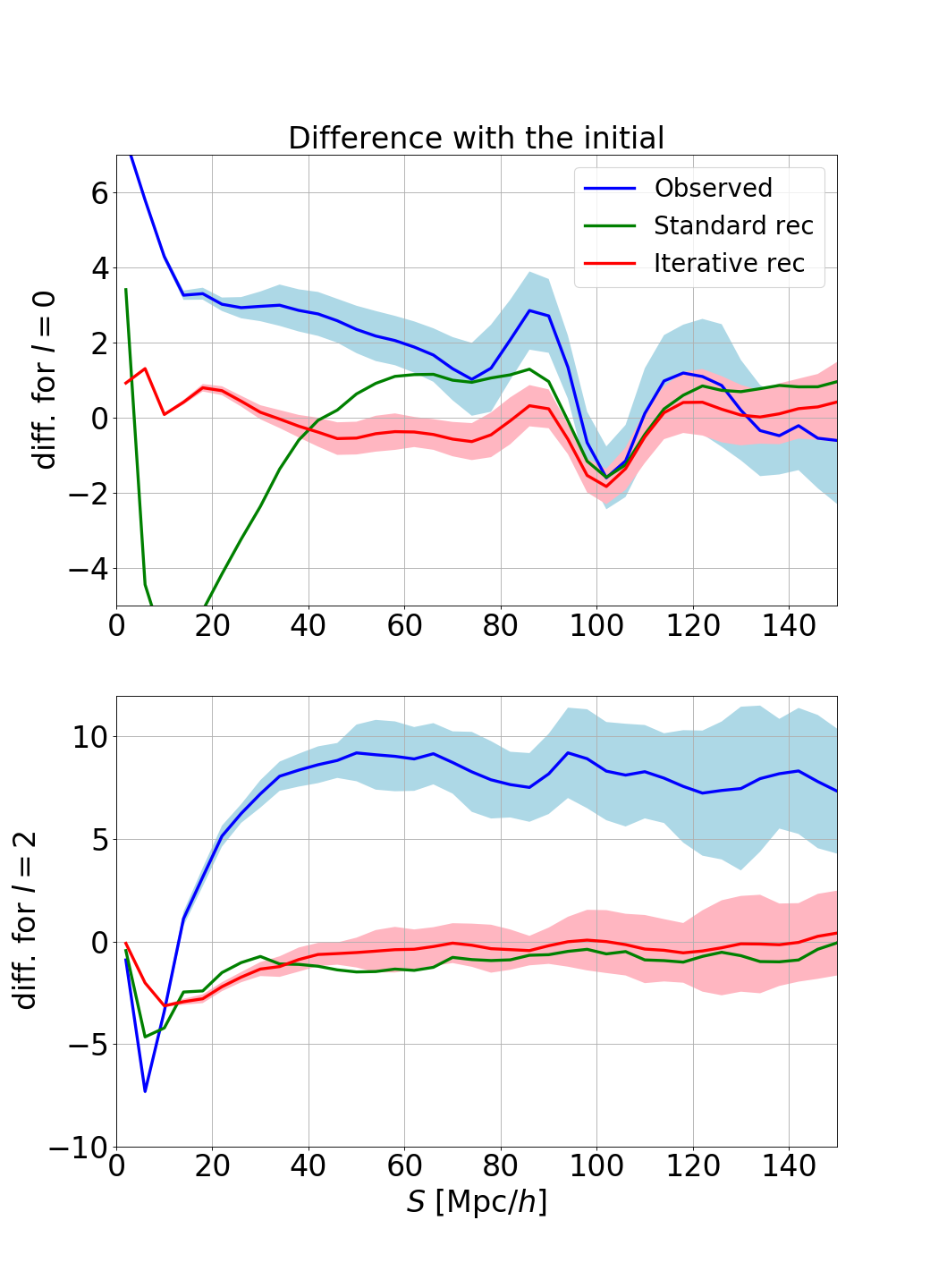}
  \end{center}
\end{minipage} \\
\begin{minipage}{0.50\hsize}
  \begin{center}
   \includegraphics[width=80mm]{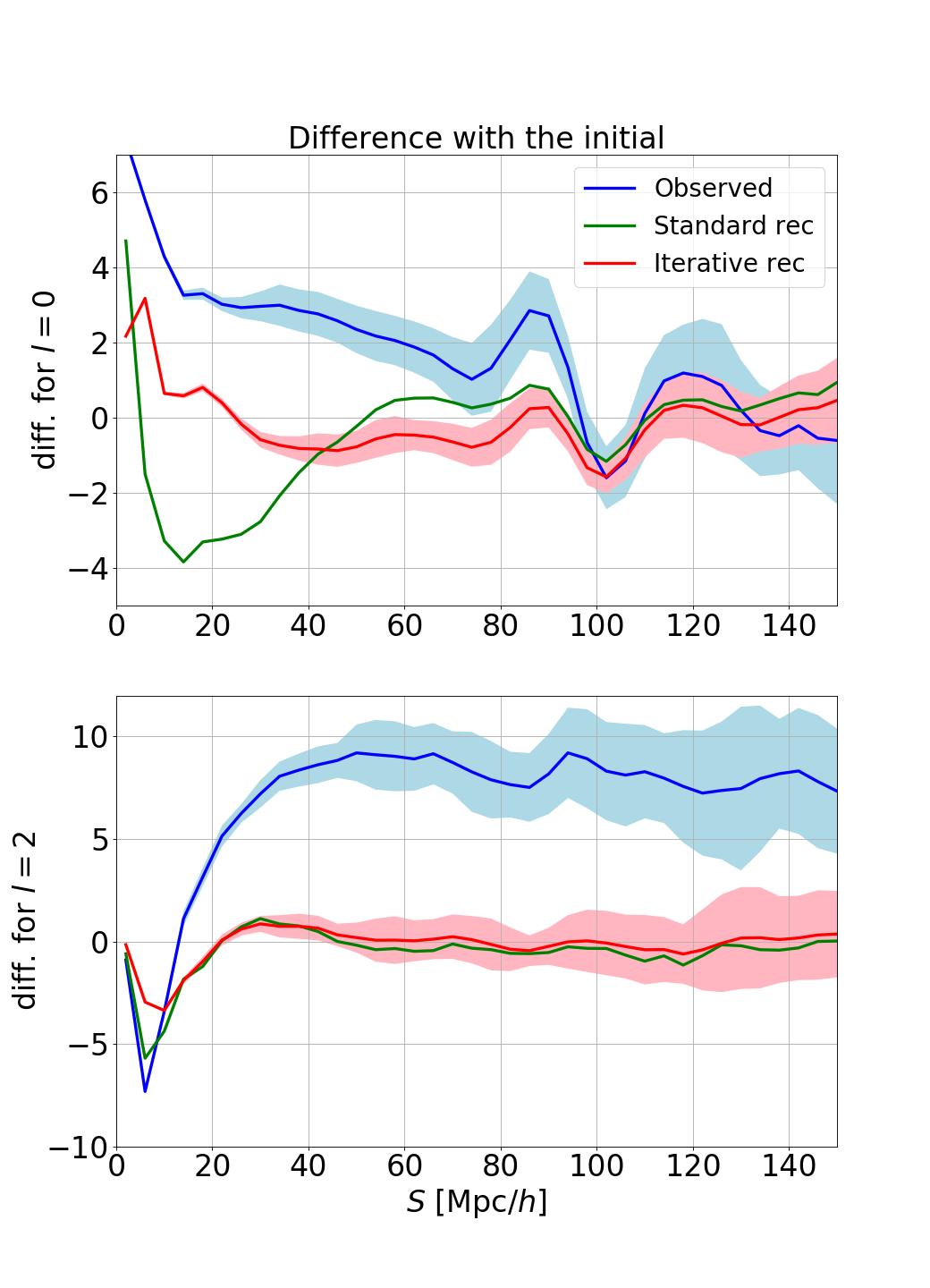}
  \end{center}
\end{minipage}
 \begin{minipage}{0.50\hsize}
  \begin{center}
   \includegraphics[width=80mm]{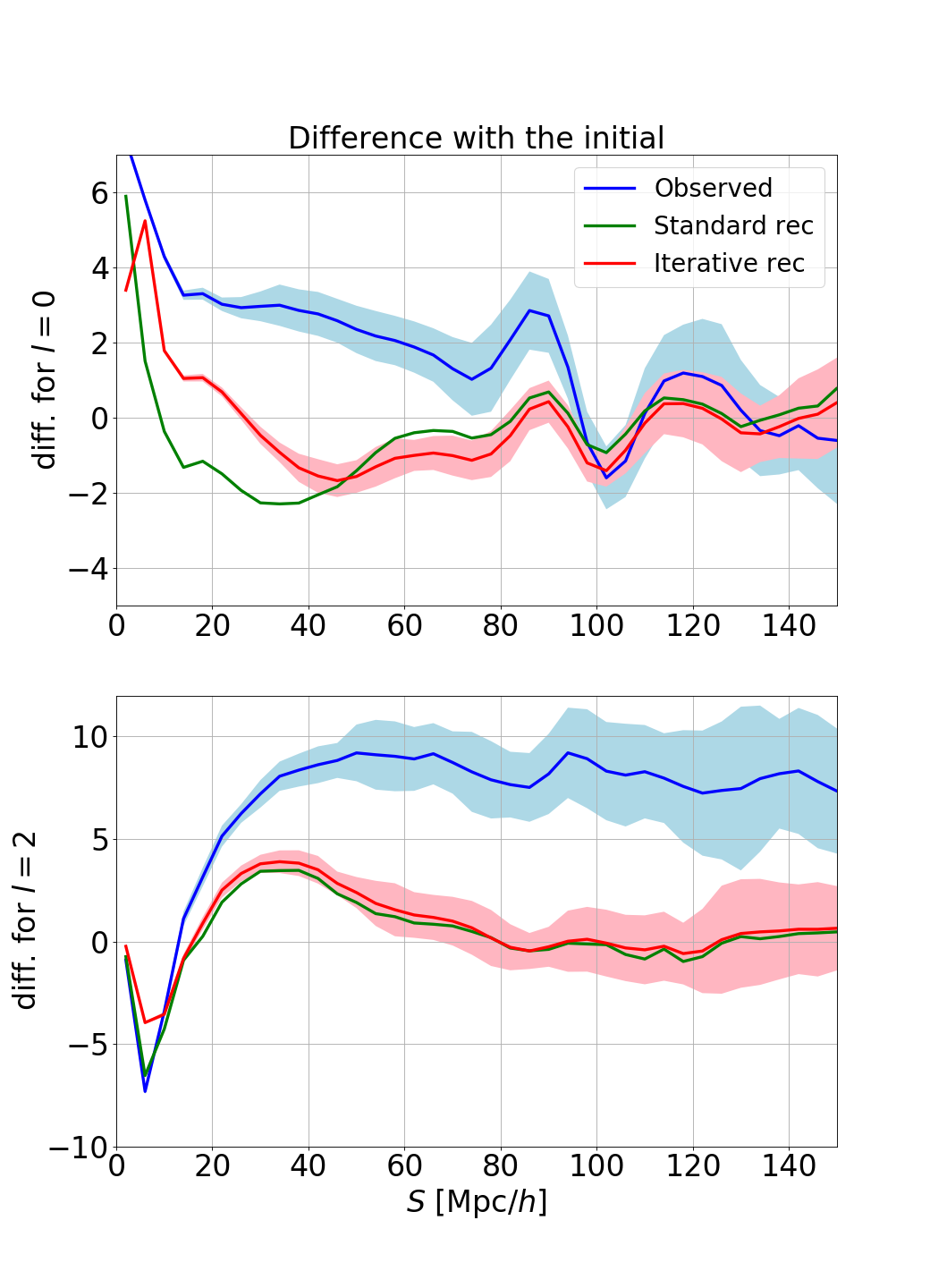}
  \end{center}
\end{minipage}
\end{tabular}
  \caption{\label{fig:sm}Comparison of smoothing scales for the galaxy fields in redshift space: $\Sigma_{\rm eff} = 5h^{-1}$ (upper left), $7h^{-1}$ (upper right), $10h^{-1}$ (lower left) and $15\hMpc$ (lower right). Each panel is described in the same manner as Fig.~\ref{fig:MvsG}. Generally, the iterative method with smaller smoothing scale can restore the monopole and quadrupole more precisely.}
\end{figure*}

\begin{figure*}
\begin{tabular}{ccc}
 \begin{minipage}{0.33\hsize}
  \begin{center}
   \includegraphics[width=65mm]{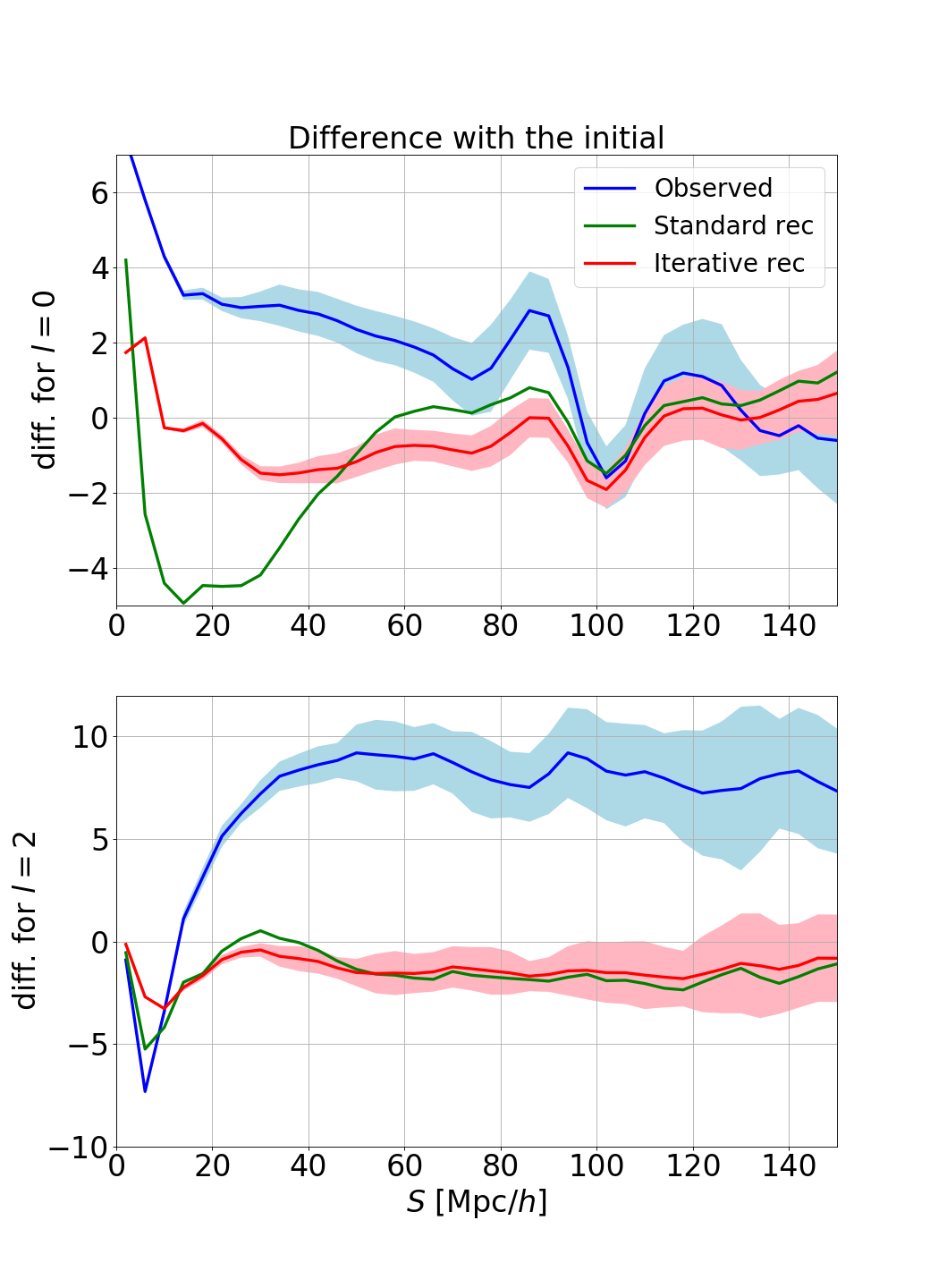}
  \end{center}
\end{minipage} 
 \begin{minipage}{0.33\hsize}
  \begin{center}
   \includegraphics[width=65mm]{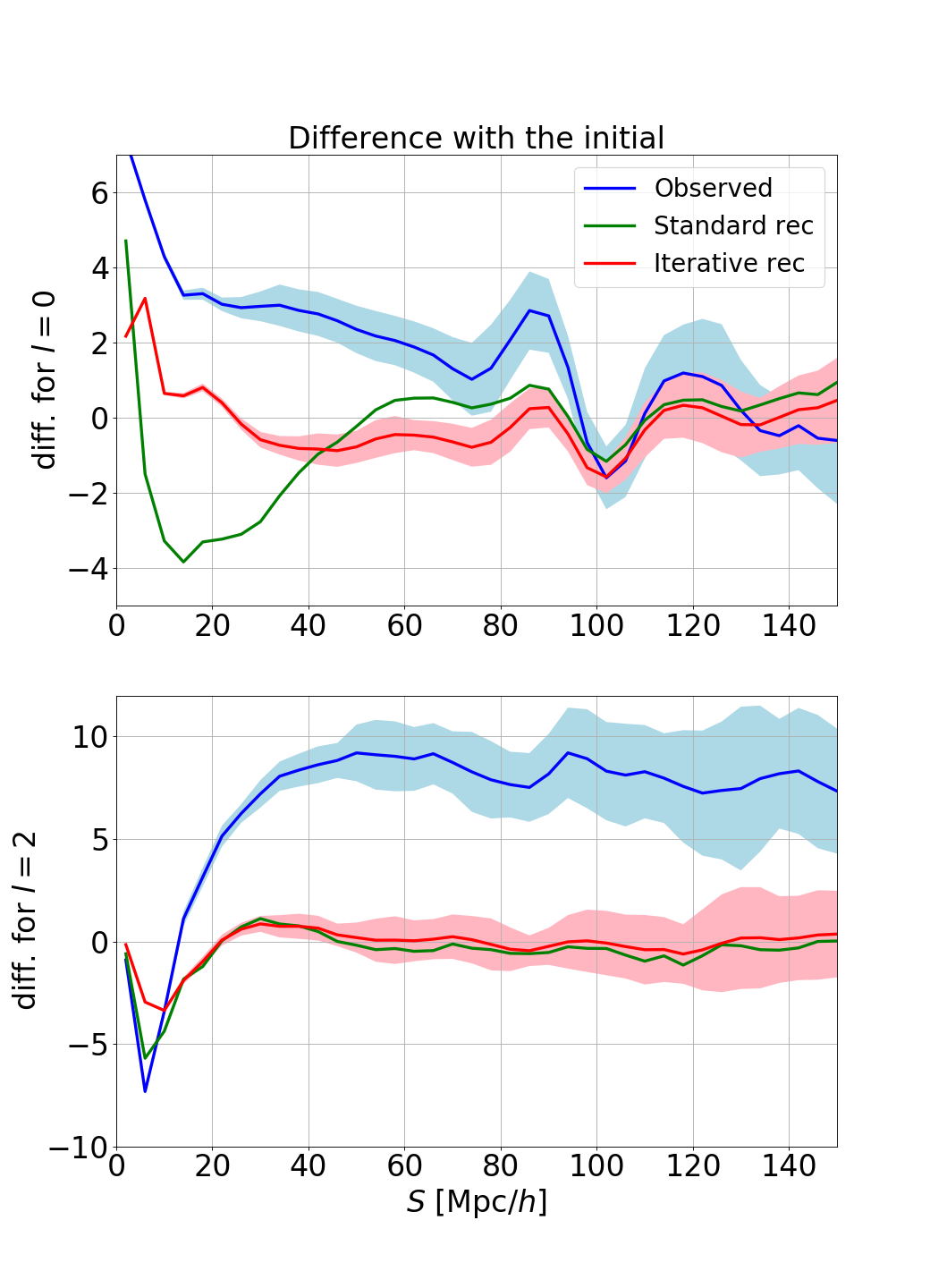}
  \end{center}
\end{minipage}
 \begin{minipage}{0.33\hsize}
  \begin{center}
   \includegraphics[width=65mm]{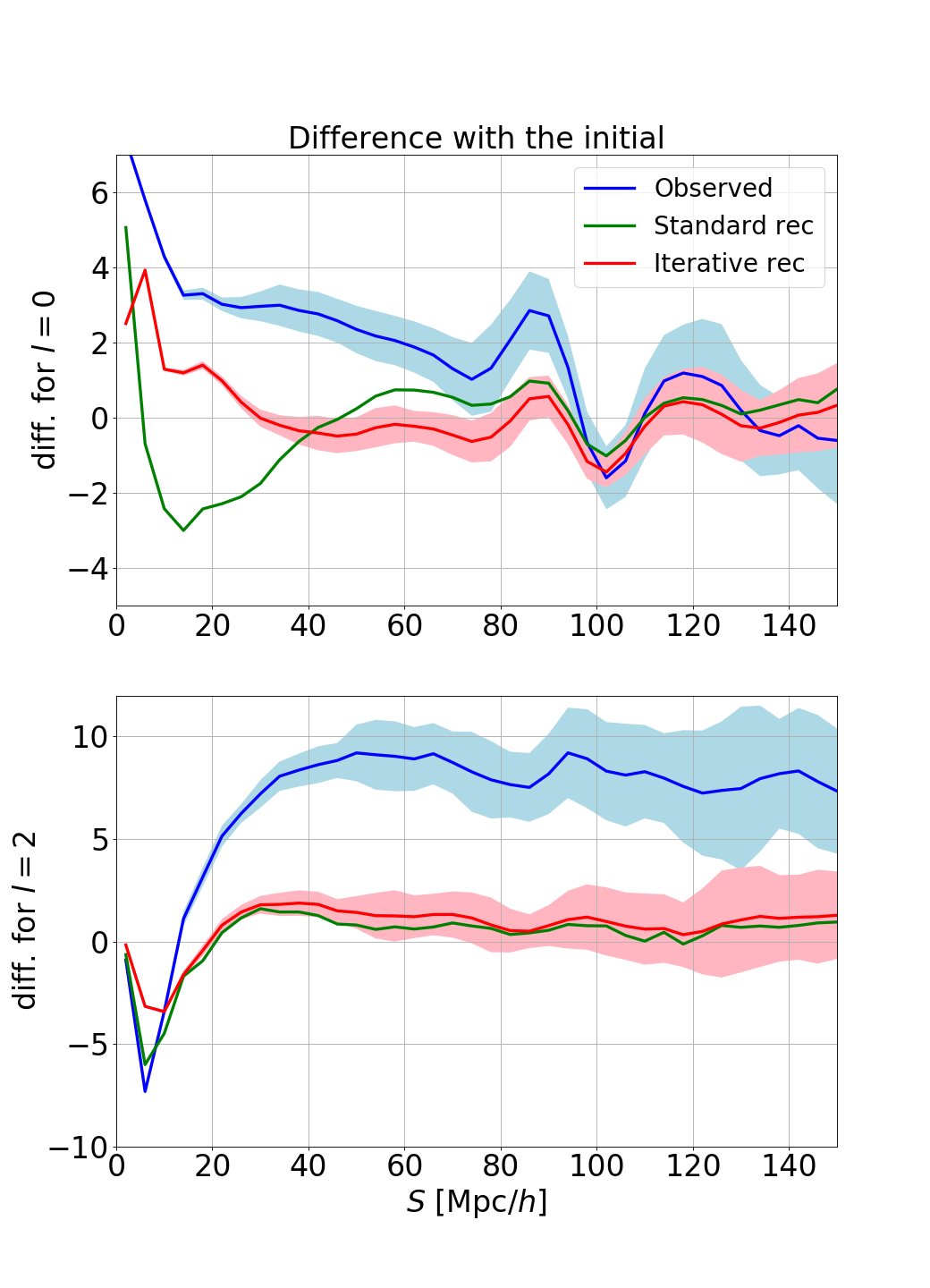}
  \end{center}
\end{minipage}
\end{tabular}
  \caption{\label{fig:bias}Comparison of the values of bias for the galaxy fields in redshift space: $b = 0.8b_{\rm f}$ (left), $b_{\rm f}$ (middle), and $1.2b_{\rm f}$ (right). Each panel is described in the same manner as Fig.~\ref{fig:MvsG}. Both reconstruction processes are hardly influenced by the change in galaxy bias.}
\end{figure*}

\begin{figure*}
\begin{tabular}{ccc}
 \begin{minipage}{0.33\hsize}
  \begin{center}
   \includegraphics[width=65mm]{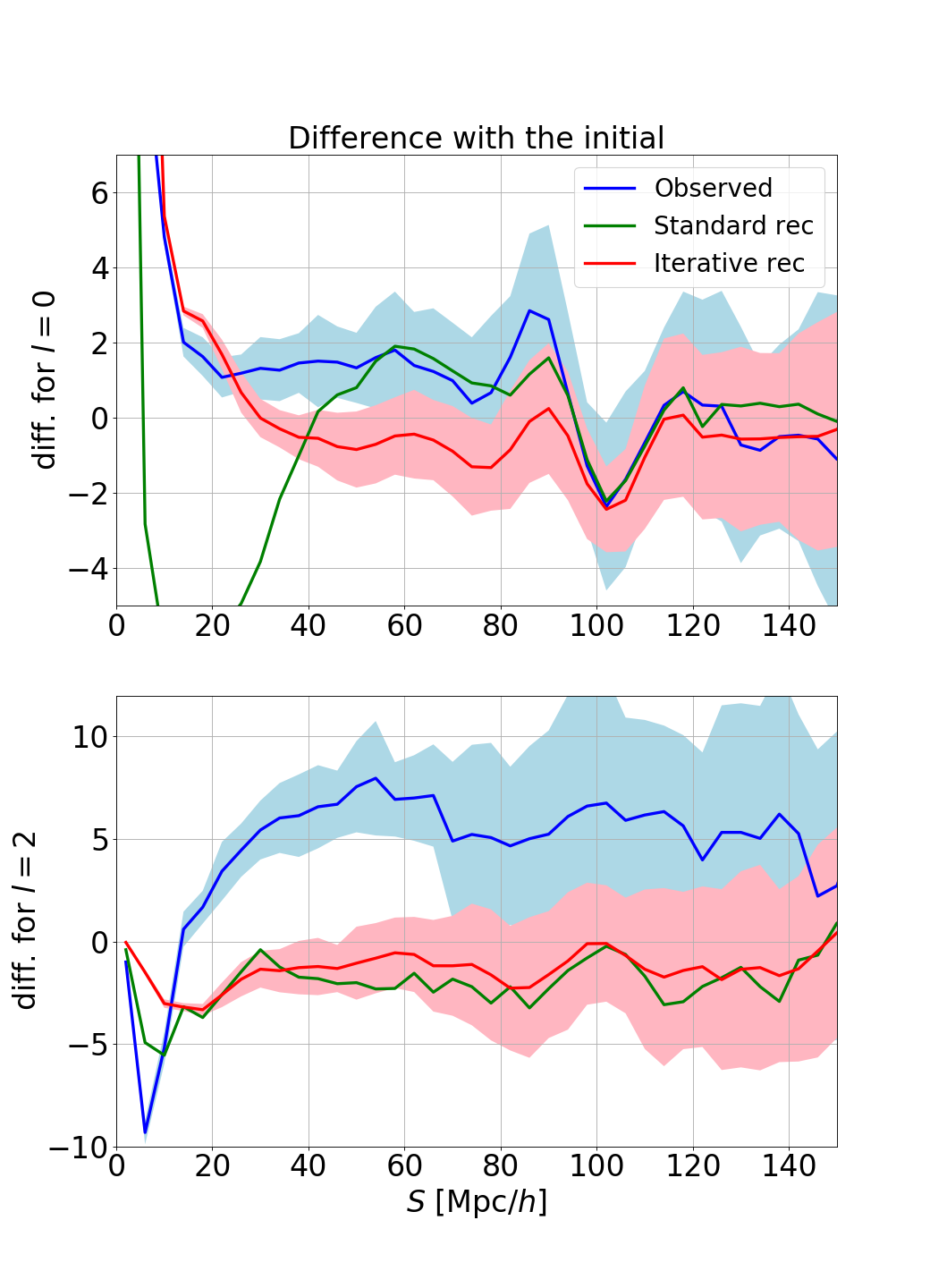}
  \end{center}
\end{minipage} 
 \begin{minipage}{0.33\hsize}
  \begin{center}
   \includegraphics[width=65mm]{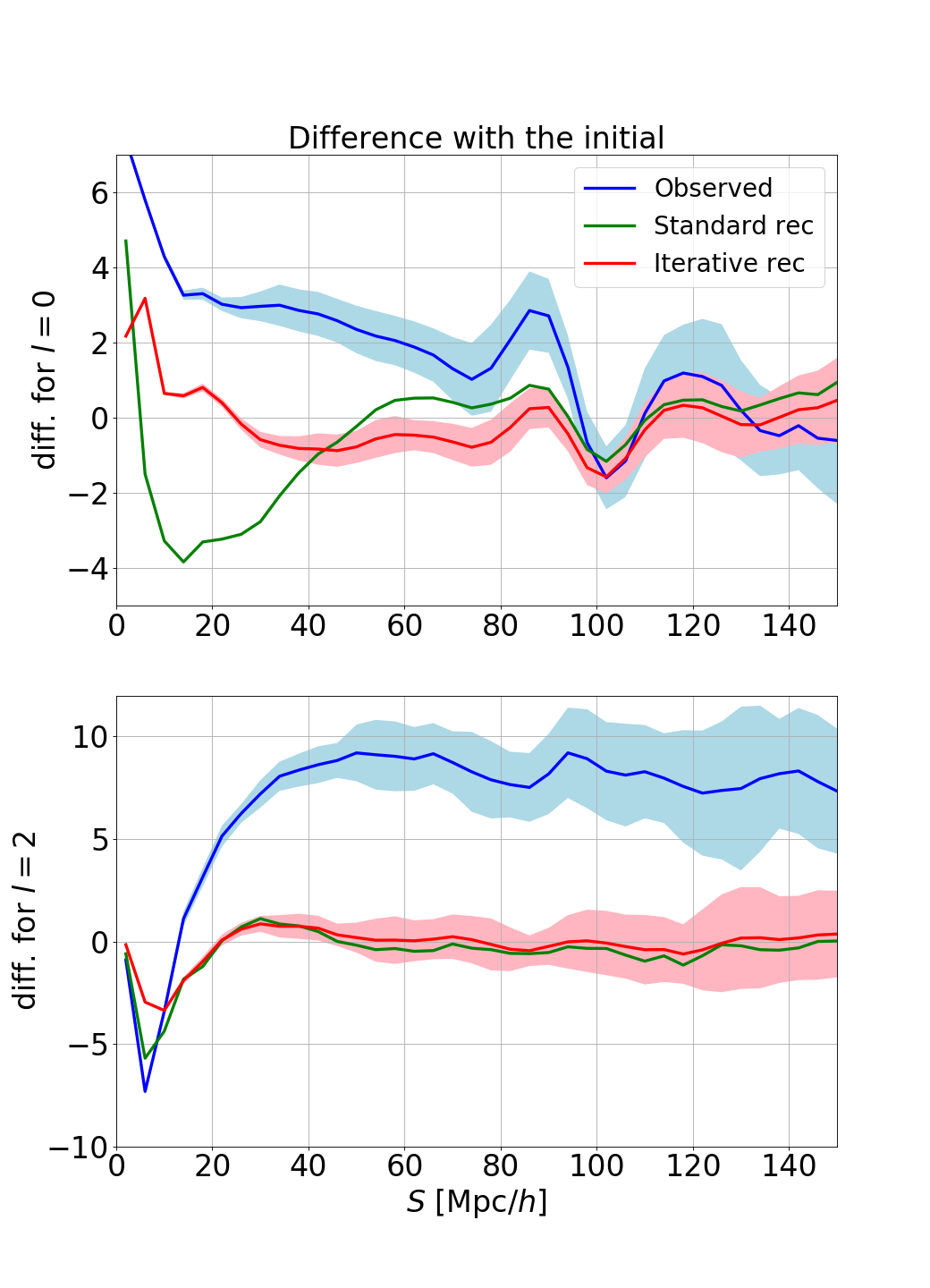}
  \end{center}
\end{minipage}
 \begin{minipage}{0.33\hsize}
  \begin{center}
   \includegraphics[width=65mm]{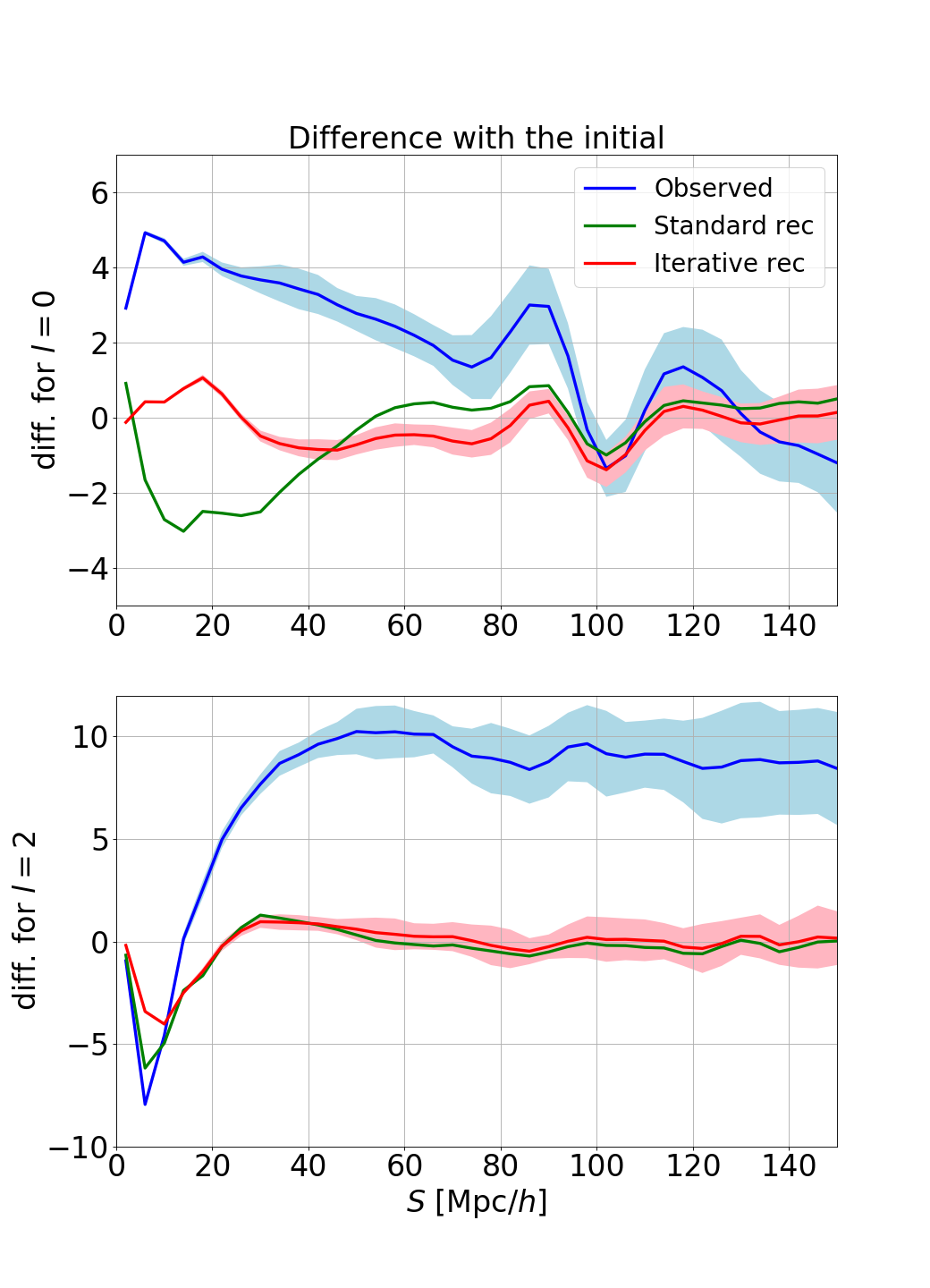}
  \end{center}
\end{minipage}
\end{tabular}
  \caption{\label{fig:N_g}Comparison of the number density of galaxies in redshift space: $n_{\rm gal} = 4.2\times10^{-5}$ ({\it smaller}, left), $4.4\times10^{-4}$ ({\it fiducial}, middle), and $2.6\times10^{-3}(\hMpc)^{-3}$ ({\it larger}, right). Each panel is described in the same manner as Fig.~\ref{fig:MvsG}. The larger the number density become, the better the performance of reconstruction gets.}
\end{figure*}

\subsubsection{\label{sec:sm}Smoothing scale}

Fig.~\ref{fig:sm} shows how the performance of reconstruction for the galaxy fields in redshift space depends on the smoothing scale $\Sigma_{\rm eff}$. First, we can see that the smaller the smoothing scale gets, the closer the monopole and quadrupole for our iterative method (in red) get to the vertical line, $\Delta \xi_{l}(S) = 0$. This reflects that the reconstruction with smaller smoothing scale can restore the displacement more precisely (see section~4.2 in \citetalias{2018MNRAS.478.1866H}). On the other hand, the performance of the standard reconstruction gets worse as the smoothing scale becomes smaller, because the effect of using the final density field instead of the initial density field computing the displacements (problem~1 in Section~\ref{sec:intro}) cannot be ignored more. 

However one find that the monopole and quadrupole in the case with $\Sigma_{\rm eff} = 5 \hMpc$ are more shifted from $\Delta \xi_{l}(S) = 0$ on small scales $S \lsim 30 \hMpc$ than the case with $\Sigma_{\rm eff} = 7 \hMpc$. We can explain about this as follows: the number of galaxies for each simulation box, $L = 1100 \hMpc$, is $\sim 84^3$, which means that the mean distance of galaxies is $\sim 13 \hMpc$. The noise in the galaxy field increases on scales smaller than the mean distance of galaxy. That is, the above results shows that $\Sigma_{\rm eff} = 5 \hMpc$ is too small to be applied to this galaxy samples (technically, we should be able to cover somewhat larger-scale modes than $\Sigma_{\rm eff}$ because of the Gaussian tail of the smoothing filter). It follows that we need to set the smoothing scale taking care of the mean distance of galaxies when applying our iterative reconstruction method to actual galaxy data.  

Furthermore, we see that as the smoothing scale increases, the differences between the iterative and standard method become smaller for both the monopole and quadrupole. This feature reflects the fact that it is harder to make the most of the advantage of making use of the iterative method (especially problem~1) when using larger smoothing scale, because the difference between the displacements estimated from the final galaxy density field and the initial linear density field can be seen only on small scales.

\subsubsection{\label{sec:bias}Bias}

In actual galaxy redshift survey, we also need to care about the uncertainty in bias estimation. To evaluate how the change in galaxy bias impacts on the reconstruction processes, we define the normalized two-point correlation as follows:
\begin{eqnarray} 
	\xi_{l, {\rm nor}}[\delta_{\rm X}](S) \equiv \left( \frac{b}{b_{\rm f}}\right)^{2} \xi_{l}[\delta_{\rm X}](S), \label{eq:xi_nor}
\end{eqnarray}
where $b_{\rm f}$ is the fiducial value of bias for each realization that is defined by Eq.~(\ref{eq:b_R}). Fig.~\ref{fig:bias} shows the differences between $\xi_{l, {\rm nor}}[\delta^{g}_{\rm L}]$ (reconstructed with $\Sigma_{\rm eff} = 10 \hMpc$) and $\xi_{l}[\delta_{\rm ini}]$ in redshift space for $b = 0.8b_{\rm f}$ (left), $b_{\rm f}$ (middle), and $1.2b_{\rm f}$ (right). As for the observed density field $\delta_{s} = \delta^{g}_{s}/b$, there is no difference among the biases because it follows from the definition above that 
\begin{eqnarray} 
	\xi_{l, {\rm nor}}[\delta^{g}_{s}/b] = \xi_{l}[\delta^{g}_{s}/b_{\rm f}]. \label{eq:xi_nor_obs}
\end{eqnarray}
Furthermore, we can hardly see the differences also for the reconstructed density fields, which means that the change of bias by $\sim20\%$ hardly impacts on the reconstruction processes. This result suggests that although we are supposed to estimate the galaxy bias from a large number of galaxy mock catalogs in actual galaxy redshift surveys, we are not bothered so much by the precision in the determination of bias when using either our iterative method or the standard method.

\subsubsection{\label{sec:N_g}Number density of galaxies}

Finally, in Fig.~\ref{fig:N_g}, we shows the results in redshift space, with different number density of galaxies: $n_{\rm gal} = 4.2\times10^{-5}$ ({\it smaller}, left), $4.4\times10^{-4}$ ({\it fiducial}, middle), and $2.6\times10^{-3}(\hMpc)^{-3}$ ({\it larger}, right). In this study, we create the galaxy fields with the smaller and larger number density by adding the factors: $0.75$ and $-0.75$, respectively, to the parameters related to halo masses, $\log M_{\rm cut}$ and $\log M_{1}$, when running GRAND-HOD. 

As the number density gets smaller, the galaxy field becomes more sparse and noisy. Indeed, we see that accordingly the variances get larger and the averages of $\Delta \xi_{l}[\delta^{g}_{\rm L}]$ (reconstructed with $\Sigma_{\rm eff} = 10 \hMpc$) shift away from $\Delta \xi_{l}(S) = 0$ entirely.

\section{Conclusions} \label{sec:Conc}

We have applied the iterative method proposed in \citetalias{2018MNRAS.478.1866H} to simulated galaxy fields in redshift space and explored some parameter choices. Focusing on the two-point correlation function in order to evaluate the performance, we found that our method can reconstruct from the galaxy field better than the standard method both in real and in redshift space, although the performance is limited by shot noise and galaxy bias compared to the matter field.

Furthermore, comparing the optimal values of $\mathcal{C}_{\rm ani}$, which manages the effect of the anisotropy in redshift space on the smoothing process, between the cases with the matter field and galaxy field, we found that the matter field is more influenced by the Finger of God effect than the galaxy field due to the effects of dynamics inside halos and smaller smoothing scale.

We also found that the iterative method with smaller smoothing scale is able to reconstruct the monopole and quadrupole more precisely unlike the standard method and that in practical cases, we need to set the smoothing scale taking account of the mean distance of galaxies. In addition, as the number density gets smaller and the galaxy bias becomes larger, the performance of reconstruction gets worse because the galaxy field becomes more sparse and noisy. On the other hand, the precision in the determination of bias ($\sim20\%$) hardly impacts on the reconstruction processes. 

In this work, we fixed the value of $f$ to a fiducial value corresponding to the fiducial cosmological parameters. The wrong assumption of $f$ (or $\beta$) might have an impact on the quadrupole, however, it is not expected to affect an acoustic scale. In addition, comparison of the reconstructed quadrupoles between calibrated simulations and data might allow us to measure $f$, and in future work we will consider whether this approach is more precise than the conventional methods that compare data with non-linear model templates without performing reconstruction.

In order to evaluate how our iterative reconstruction method actually improves the distance measurement, we need to evaluate the precision in the BAO distance measurement by fitting the acoustic signature to a template. Our iterative method shows the better accuracy than the standard method not only on small scales, but also on intermediate scales $40 \lsim S \lsim 90 \hMpc$. Therefore it is expected to improve the fitting procedure. We will defer such investigation to future work.

Regardless, the iterative method has the advantage of being able to solve the problems that the standard method has (see Section~\ref{sec:intro}). In particular, considering that real surveys have boundaries, the merit of being able to avoid displacing data and random particles (problem~4) should be effective in the upcoming galaxy surveys where the survey density is rapidly varying in the radial direction. Thus we expect that the iterative reconstruction method can make the BAO standard ruler more reliable in upcoming surveys.

\section*{Acknowledgements}

We would like to thank Lehman H. Garrison and Sihan Yuan for useful discussions. RH is supported by Japan Society for the Promotion of Science, Research Fellowships for Young Scientists (No. 16J01773) and as a doctoral course student in the Division for Interdisciplinary Advanced Research and Education, Tohoku University. DJE is supported by U.S.\ Department of Energy grant DE-SC0013718 and as a Simons Foundation Investigator.

\bibliographystyle{mnras}
\bibliography{ms}

\begin{thebibliography}{}
\makeatletter
\relax
\def\mn@urlcharsother{\let\do\@makeother \do\$\do\&\do\#\do\^\do\_\do\%\do\~}
\def\mn@doi{\begingroup\mn@urlcharsother \@ifnextchar [ {\mn@doi@}
  {\mn@doi@[]}}
\def\mn@doi@[#1]#2{\def\@tempa{#1}\ifx\@tempa\@empty \href
  {http://dx.doi.org/#2} {doi:#2}\else \href {http://dx.doi.org/#2} {#1}\fi
  \endgroup}
\def\mn@eprint#1#2{\mn@eprint@#1:#2::\@nil}
\def\mn@eprint@arXiv#1{\href {http://arxiv.org/abs/#1} {{\tt arXiv:#1}}}
\def\mn@eprint@dblp#1{\href {http://dblp.uni-trier.de/rec/bibtex/#1.xml}
  {dblp:#1}}
\def\mn@eprint@#1:#2:#3:#4\@nil{\def\@tempa {#1}\def\@tempb {#2}\def\@tempc
  {#3}\ifx \@tempc \@empty \let \@tempc \@tempb \let \@tempb \@tempa \fi \ifx
  \@tempb \@empty \def\@tempb {arXiv}\fi \@ifundefined
  {mn@eprint@\@tempb}{\@tempb:\@tempc}{\expandafter \expandafter \csname
  mn@eprint@\@tempb\endcsname \expandafter{\@tempc}}}

\bibitem[\protect\citeauthoryear{{Anderson} et~al.,}{{Anderson}
  et~al.}{2012}]{2012MNRAS.427.3435A}
{Anderson} L.,  et~al., 2012, \mn@doi [\mnras]
  {10.1111/j.1365-2966.2012.22066.x}, \href
  {http://adsabs.harvard.edu/abs/2012MNRAS.427.3435A} {427, 3435}

\bibitem[\protect\citeauthoryear{{Anderson} et~al.,}{{Anderson}
  et~al.}{2014}]{2014MNRAS.441...24A}
{Anderson} L.,  et~al., 2014, \mn@doi [\mnras] {10.1093/mnras/stu523}, \href
  {http://adsabs.harvard.edu/abs/2014MNRAS.441...24A} {441, 24}

\bibitem[\protect\citeauthoryear{{Angulo}, {Baugh}, {Frenk}, {Bower}, {Jenkins}
   \& {Morris}}{{Angulo} et~al.}{2005}]{2005MNRAS.362L..25A}
{Angulo} R.~E.,  {Baugh} C.~M.,  {Frenk} C.~S.,  {Bower} R.~G.,  {Jenkins} A.,
   {Morris} S.~L.,  2005, \mn@doi [\mnras] {10.1111/j.1745-3933.2005.00067.x},
  \href {http://adsabs.harvard.edu/abs/2005MNRAS.362L..25A} {362, L25}

\bibitem[\protect\citeauthoryear{{Angulo}, {Baugh}, {Frenk}  \&
  {Lacey}}{{Angulo} et~al.}{2008}]{2008MNRAS.383..755A}
{Angulo} R.~E.,  {Baugh} C.~M.,  {Frenk} C.~S.,   {Lacey} C.~G.,  2008, \mn@doi
  [\mnras] {10.1111/j.1365-2966.2007.12587.x}, \href
  {http://adsabs.harvard.edu/abs/2008MNRAS.383..755A} {383, 755}

\bibitem[\protect\citeauthoryear{{Behroozi}, {Wechsler}  \& {Wu}}{{Behroozi}
  et~al.}{2013}]{2013ApJ...762..109B}
{Behroozi} P.~S.,  {Wechsler} R.~H.,   {Wu} H.-Y.,  2013, \mn@doi [\apj]
  {10.1088/0004-637X/762/2/109}, \href
  {http://adsabs.harvard.edu/abs/2013ApJ...762..109B} {762, 109}

\bibitem[\protect\citeauthoryear{{Bernardeau}, {Colombi}, {Gazta{\~n}aga}  \&
  {Scoccimarro}}{{Bernardeau} et~al.}{2002}]{2002PhR...367....1B}
{Bernardeau} F.,  {Colombi} S.,  {Gazta{\~n}aga} E.,   {Scoccimarro} R.,  2002,
  \mn@doi [\physrep] {10.1016/S0370-1573(02)00135-7}, \href
  {http://adsabs.harvard.edu/abs/2002PhR...367....1B} {367, 1}

\bibitem[\protect\citeauthoryear{{Burden}, {Percival}, {Manera}, {Cuesta},
  {Vargas Magana}  \& {Ho}}{{Burden} et~al.}{2014}]{2014MNRAS.445.3152B}
{Burden} A.,  {Percival} W.~J.,  {Manera} M.,  {Cuesta} A.~J.,  {Vargas Magana}
  M.,   {Ho} S.,  2014, \mn@doi [\mnras] {10.1093/mnras/stu1965}, \href
  {http://adsabs.harvard.edu/abs/2014MNRAS.445.3152B} {445, 3152}

\bibitem[\protect\citeauthoryear{{Cole} et~al.,}{{Cole}
  et~al.}{2005}]{2005MNRAS.362..505C}
{Cole} S.,  et~al., 2005, \mn@doi [\mnras] {10.1111/j.1365-2966.2005.09318.x},
  \href {http://adsabs.harvard.edu/abs/2005MNRAS.362..505C} {362, 505}

\bibitem[\protect\citeauthoryear{{DESI Collaboration} et~al.,}{{DESI
  Collaboration} et~al.}{2016}]{2016arXiv161100036D}
{DESI Collaboration} et~al., 2016, preprint, \href
  {http://adsabs.harvard.edu/abs/2016arXiv161100036D} {} (\mn@eprint {arXiv}
  {1611.00036})

\bibitem[\protect\citeauthoryear{{Eisenstein}, {Blanton}, {Zehavi}, {Bahcall},
  {Brinkmann}, {Loveday}, {Meiksin}  \& {Schneider}}{{Eisenstein}
  et~al.}{2005a}]{2005ApJ...619..178E}
{Eisenstein} D.~J.,  {Blanton} M.,  {Zehavi} I.,  {Bahcall} N.,  {Brinkmann}
  J.,  {Loveday} J.,  {Meiksin} A.,   {Schneider} D.,  2005a, \mn@doi [\apj]
  {10.1086/426500}, \href {http://adsabs.harvard.edu/abs/2005ApJ...619..178E}
  {619, 178}

\bibitem[\protect\citeauthoryear{{Eisenstein} et~al.,}{{Eisenstein}
  et~al.}{2005b}]{2005ApJ...633..560E}
{Eisenstein} D.~J.,  et~al., 2005b, \mn@doi [\apj] {10.1086/466512}, \href
  {http://adsabs.harvard.edu/abs/2005ApJ...633..560E} {633, 560}

\bibitem[\protect\citeauthoryear{{Eisenstein}, {Seo}  \& {White}}{{Eisenstein}
  et~al.}{2007a}]{2007ApJ...664..660E}
{Eisenstein} D.~J.,  {Seo} H.-J.,   {White} M.,  2007a, \mn@doi [\apj]
  {10.1086/518755}, \href {http://adsabs.harvard.edu/abs/2007ApJ...664..660E}
  {664, 660}

\bibitem[\protect\citeauthoryear{{Eisenstein}, {Seo}, {Sirko}  \&
  {Spergel}}{{Eisenstein} et~al.}{2007b}]{2007ApJ...664..675E}
{Eisenstein} D.~J.,  {Seo} H.-J.,  {Sirko} E.,   {Spergel} D.~N.,  2007b,
  \mn@doi [\apj] {10.1086/518712}, \href
  {http://adsabs.harvard.edu/abs/2007ApJ...664..675E} {664, 675}

\bibitem[\protect\citeauthoryear{{Garrison}, {Eisenstein}, {Ferrer}, {Metchnik}
   \& {Pinto}}{{Garrison} et~al.}{2016}]{2016MNRAS.461.4125G}
{Garrison} L.~H.,  {Eisenstein} D.~J.,  {Ferrer} D.,  {Metchnik} M.~V.,
  {Pinto} P.~A.,  2016, \mn@doi [\mnras] {10.1093/mnras/stw1594}, \href
  {http://adsabs.harvard.edu/abs/2016MNRAS.461.4125G} {461, 4125}

\bibitem[\protect\citeauthoryear{{Garrison}, {Eisenstein}, {Ferrer}, {Tinker},
  {Pinto}  \& {Weinberg}}{{Garrison} et~al.}{2017}]{2017arXiv171205768G}
{Garrison} L.~H.,  {Eisenstein} D.~J.,  {Ferrer} D.,  {Tinker} J.~L.,  {Pinto}
  P.~A.,   {Weinberg} D.~H.,  2017, preprint, \href
  {http://adsabs.harvard.edu/abs/2017arXiv171205768G} {} (\mn@eprint {arXiv}
  {1712.05768})

\bibitem[\protect\citeauthoryear{{Garrison}, {Eisenstein}  \&
  {Pinto}}{{Garrison} et~al.}{2018}]{2018arXiv181002916G}
{Garrison} L.~H.,  {Eisenstein} D.~J.,   {Pinto} P.~A.,  2018, preprint, \href
  {http://adsabs.harvard.edu/abs/2018arXiv181002916G} {} (\mn@eprint {arXiv}
  {1810.02916})

\bibitem[\protect\citeauthoryear{{Hada} \& {Eisenstein}}{{Hada} \&
  {Eisenstein}}{2018}]{2018MNRAS.478.1866H}
{Hada} R.,  {Eisenstein} D.~J.,  2018, \mn@doi [\mnras]
  {10.1093/mnras/sty1203}, \href
  {http://adsabs.harvard.edu/abs/2018MNRAS.478.1866H} {478, 1866}

\bibitem[\protect\citeauthoryear{{Huff}, {Schulz}, {White}, {Schlegel}  \&
  {Warren}}{{Huff} et~al.}{2007}]{2007APh....26..351H}
{Huff} E.,  {Schulz} A.~E.,  {White} M.,  {Schlegel} D.~J.,   {Warren} M.~S.,
  2007, \mn@doi [Astroparticle Physics] {10.1016/j.astropartphys.2006.07.007},
  \href {http://adsabs.harvard.edu/abs/2007APh....26..351H} {26, 351}

\bibitem[\protect\citeauthoryear{{Jeong} \& {Komatsu}}{{Jeong} \&
  {Komatsu}}{2006}]{2006ApJ...651..619J}
{Jeong} D.,  {Komatsu} E.,  2006, \mn@doi [\apj] {10.1086/507781}, \href
  {http://adsabs.harvard.edu/abs/2006ApJ...651..619J} {651, 619}

\bibitem[\protect\citeauthoryear{{Kaiser}}{{Kaiser}}{1987}]{1987MNRAS.227....1K}
{Kaiser} N.,  1987, \mn@doi [\mnras] {10.1093/mnras/227.1.1}, \href
  {http://adsabs.harvard.edu/abs/1987MNRAS.227....1K} {227, 1}

\bibitem[\protect\citeauthoryear{{Kwan}, {Heitmann}, {Habib}, {Padmanabhan},
  {Lawrence}, {Finkel}, {Frontiere}  \& {Pope}}{{Kwan}
  et~al.}{2015}]{2015ApJ...810...35K}
{Kwan} J.,  {Heitmann} K.,  {Habib} S.,  {Padmanabhan} N.,  {Lawrence} E.,
  {Finkel} H.,  {Frontiere} N.,   {Pope} A.,  2015, \mn@doi [\apj]
  {10.1088/0004-637X/810/1/35}, \href
  {http://adsabs.harvard.edu/abs/2015ApJ...810...35K} {810, 35}

\bibitem[\protect\citeauthoryear{{Laureijs} et~al.,}{{Laureijs}
  et~al.}{2011}]{2011arXiv1110.3193L}
{Laureijs} R.,  et~al., 2011, preprint, \href
  {http://adsabs.harvard.edu/abs/2011arXiv1110.3193L} {} (\mn@eprint {arXiv}
  {1110.3193})

\bibitem[\protect\citeauthoryear{{Mehta}, {Seo}, {Eckel}, {Eisenstein},
  {Metchnik}, {Pinto}  \& {Xu}}{{Mehta} et~al.}{2011}]{2011ApJ...734...94M}
{Mehta} K.~T.,  {Seo} H.-J.,  {Eckel} J.,  {Eisenstein} D.~J.,  {Metchnik} M.,
  {Pinto} P.,   {Xu} X.,  2011, \mn@doi [\apj] {10.1088/0004-637X/734/2/94},
  \href {http://adsabs.harvard.edu/abs/2011ApJ...734...94M} {734, 94}

\bibitem[\protect\citeauthoryear{{Meiksin}, {White}  \& {Peacock}}{{Meiksin}
  et~al.}{1999}]{1999MNRAS.304..851M}
{Meiksin} A.,  {White} M.,   {Peacock} J.~A.,  1999, \mn@doi [\mnras]
  {10.1046/j.1365-8711.1999.02369.x}, \href
  {http://adsabs.harvard.edu/abs/1999MNRAS.304..851M} {304, 851}

\bibitem[\protect\citeauthoryear{{Monaco} \& {Efstathiou}}{{Monaco} \&
  {Efstathiou}}{1999}]{1999MNRAS.308..763M}
{Monaco} P.,  {Efstathiou} G.,  1999, \mn@doi [\mnras]
  {10.1046/j.1365-8711.1999.02747.x}, \href
  {http://adsabs.harvard.edu/abs/1999MNRAS.308..763M} {308, 763}

\bibitem[\protect\citeauthoryear{{Noh}, {White}  \& {Padmanabhan}}{{Noh}
  et~al.}{2009}]{2009PhRvD..80l3501N}
{Noh} Y.,  {White} M.,   {Padmanabhan} N.,  2009, \mn@doi [\prd]
  {10.1103/PhysRevD.80.123501}, \href
  {http://adsabs.harvard.edu/abs/2009PhRvD..80l3501N} {80, 123501}

\bibitem[\protect\citeauthoryear{{Padmanabhan}, {Xu}, {Eisenstein}, {Scalzo},
  {Cuesta}, {Mehta}  \& {Kazin}}{{Padmanabhan}
  et~al.}{2012}]{2012MNRAS.427.2132P}
{Padmanabhan} N.,  {Xu} X.,  {Eisenstein} D.~J.,  {Scalzo} R.,  {Cuesta} A.~J.,
   {Mehta} K.~T.,   {Kazin} E.,  2012, \mn@doi [\mnras]
  {10.1111/j.1365-2966.2012.21888.x}, \href
  {http://adsabs.harvard.edu/abs/2012MNRAS.427.2132P} {427, 2132}

\bibitem[\protect\citeauthoryear{{Peebles} \& {Yu}}{{Peebles} \&
  {Yu}}{1970}]{1970ApJ...162..815P}
{Peebles} P.~J.~E.,  {Yu} J.~T.,  1970, \mn@doi [\apj] {10.1086/150713}, \href
  {http://adsabs.harvard.edu/abs/1970ApJ...162..815P} {162, 815}

\bibitem[\protect\citeauthoryear{{Planck Collaboration} et~al.,}{{Planck
  Collaboration} et~al.}{2016}]{2016A&A...594A..13P}
{Planck Collaboration} et~al., 2016, \mn@doi [\aap]
  {10.1051/0004-6361/201525830}, \href
  {http://adsabs.harvard.edu/abs/2016A%26A...594A..13P} {594, A13}

\bibitem[\protect\citeauthoryear{{Ross} et~al.,}{{Ross}
  et~al.}{2017}]{2017MNRAS.464.1168R}
{Ross} A.~J.,  et~al., 2017, \mn@doi [\mnras] {10.1093/mnras/stw2372}, \href
  {http://adsabs.harvard.edu/abs/2017MNRAS.464.1168R} {464, 1168}

\bibitem[\protect\citeauthoryear{{Schmittfull}, {Baldauf}  \&
  {Zaldarriaga}}{{Schmittfull} et~al.}{2017}]{2017PhRvD..96b3505S}
{Schmittfull} M.,  {Baldauf} T.,   {Zaldarriaga} M.,  2017, \mn@doi [\prd]
  {10.1103/PhysRevD.96.023505}, \href
  {http://adsabs.harvard.edu/abs/2017PhRvD..96b3505S} {96, 023505}

\bibitem[\protect\citeauthoryear{{Seo} \& {Eisenstein}}{{Seo} \&
  {Eisenstein}}{2005}]{2005ApJ...633..575S}
{Seo} H.-J.,  {Eisenstein} D.~J.,  2005, \mn@doi [\apj] {10.1086/491599}, \href
  {http://adsabs.harvard.edu/abs/2005ApJ...633..575S} {633, 575}

\bibitem[\protect\citeauthoryear{{Seo} et~al.,}{{Seo}
  et~al.}{2010}]{2010ApJ...720.1650S}
{Seo} H.-J.,  et~al., 2010, \mn@doi [\apj] {10.1088/0004-637X/720/2/1650},
  \href {http://adsabs.harvard.edu/abs/2010ApJ...720.1650S} {720, 1650}

\bibitem[\protect\citeauthoryear{{Shi}, {Cautun}  \& {Li}}{{Shi}
  et~al.}{2018}]{2018PhRvD..97b3505S}
{Shi} Y.,  {Cautun} M.,   {Li} B.,  2018, \mn@doi [\prd]
  {10.1103/PhysRevD.97.023505}, \href
  {http://adsabs.harvard.edu/abs/2018PhRvD..97b3505S} {97, 023505}

\bibitem[\protect\citeauthoryear{{Springel} et~al.,}{{Springel}
  et~al.}{2005}]{2005Natur.435..629S}
{Springel} V.,  et~al., 2005, \mn@doi [\nat] {10.1038/nature03597}, \href
  {http://adsabs.harvard.edu/abs/2005Natur.435..629S} {435, 629}

\bibitem[\protect\citeauthoryear{{Sunyaev} \& {Zeldovich}}{{Sunyaev} \&
  {Zeldovich}}{1970}]{1970Ap&SS...7....3S}
{Sunyaev} R.~A.,  {Zeldovich} Y.~B.,  1970, \mn@doi [\apss]
  {10.1007/BF00653471}, \href
  {http://adsabs.harvard.edu/abs/1970Ap%26SS...7....3S} {7, 3}

\bibitem[\protect\citeauthoryear{{Takada} et~al.,}{{Takada}
  et~al.}{2014}]{2014PASJ...66R...1T}
{Takada} M.,  et~al., 2014, \mn@doi [\pasj] {10.1093/pasj/pst019}, \href
  {http://adsabs.harvard.edu/abs/2014PASJ...66R...1T} {66, R1}

\bibitem[\protect\citeauthoryear{{Tassev} \& {Zaldarriaga}}{{Tassev} \&
  {Zaldarriaga}}{2012}]{2012JCAP...10..006T}
{Tassev} S.,  {Zaldarriaga} M.,  2012, \mn@doi [\jcap]
  {10.1088/1475-7516/2012/10/006}, \href
  {http://adsabs.harvard.edu/abs/2012JCAP...10..006T} {10, 006}

\bibitem[\protect\citeauthoryear{{Vargas-Maga{\~n}a}
  et~al.,}{{Vargas-Maga{\~n}a} et~al.}{2016}]{2016arXiv161003506V}
{Vargas-Maga{\~n}a} M.,  et~al., 2016, preprint, \href
  {http://adsabs.harvard.edu/abs/2016arXiv161003506V} {} (\mn@eprint {arXiv}
  {1610.03506})

\bibitem[\protect\citeauthoryear{{Wang} \& {Pen}}{{Wang} \&
  {Pen}}{2018}]{2018arXiv180706381W}
{Wang} X.,  {Pen} U.-L.,  2018, preprint, \href
  {http://adsabs.harvard.edu/abs/2018arXiv180706381W} {} (\mn@eprint {arXiv}
  {1807.06381})

\bibitem[\protect\citeauthoryear{{Weinberg}, {Mortonson}, {Eisenstein},
  {Hirata}, {Riess}  \& {Rozo}}{{Weinberg} et~al.}{2013}]{2013PhR...530...87W}
{Weinberg} D.~H.,  {Mortonson} M.~J.,  {Eisenstein} D.~J.,  {Hirata} C.,
  {Riess} A.~G.,   {Rozo} E.,  2013, \mn@doi [\physrep]
  {10.1016/j.physrep.2013.05.001}, \href
  {http://adsabs.harvard.edu/abs/2013PhR...530...87W} {530, 87}

\bibitem[\protect\citeauthoryear{{Yu}, {Zhu}  \& {Pen}}{{Yu}
  et~al.}{2017}]{2017ApJ...847..110Y}
{Yu} Y.,  {Zhu} H.-M.,   {Pen} U.-L.,  2017, \mn@doi [\apj]
  {10.3847/1538-4357/aa89e7}, \href
  {http://adsabs.harvard.edu/abs/2017ApJ...847..110Y} {847, 110}

\bibitem[\protect\citeauthoryear{{Yuan}, {Eisenstein}  \& {Garrison}}{{Yuan}
  et~al.}{2018}]{2018MNRAS.tmp.1043Y}
{Yuan} S.,  {Eisenstein} D.~J.,   {Garrison} L.~H.,  2018, \mn@doi [\mnras]
  {10.1093/mnras/sty1089}, \href
  {http://adsabs.harvard.edu/abs/2018MNRAS.tmp.1043Y} {}

\bibitem[\protect\citeauthoryear{{Zehavi} et~al.,}{{Zehavi}
  et~al.}{2005}]{2005ApJ...621...22Z}
{Zehavi} I.,  et~al., 2005, \mn@doi [\apj] {10.1086/427495}, \href
  {http://adsabs.harvard.edu/abs/2005ApJ...621...22Z} {621, 22}

\bibitem[\protect\citeauthoryear{{Zel'dovich}}{{Zel'dovich}}{1970}]{1970A&A.....5...84Z}
{Zel'dovich} Y.~B.,  1970, \aap, \href
  {http://adsabs.harvard.edu/abs/1970A%26A.....5...84Z} {5, 84}

\bibitem[\protect\citeauthoryear{{Zheng}, {Zehavi}, {Eisenstein}, {Weinberg}
  \& {Jing}}{{Zheng} et~al.}{2009}]{2009ApJ...707..554Z}
{Zheng} Z.,  {Zehavi} I.,  {Eisenstein} D.~J.,  {Weinberg} D.~H.,   {Jing}
  Y.~P.,  2009, \mn@doi [\apj] {10.1088/0004-637X/707/1/554}, \href
  {http://adsabs.harvard.edu/abs/2009ApJ...707..554Z} {707, 554}

\bibitem[\protect\citeauthoryear{{Zhu}, {Yu}, {Pen}, {Chen}  \& {Yu}}{{Zhu}
  et~al.}{2017}]{2017PhRvD..96l3502Z}
{Zhu} H.-M.,  {Yu} Y.,  {Pen} U.-L.,  {Chen} X.,   {Yu} H.-R.,  2017, \mn@doi
  [\prd] {10.1103/PhysRevD.96.123502}, \href
  {http://adsabs.harvard.edu/abs/2017PhRvD..96l3502Z} {96, 123502}

\makeatother
\end{thebibliography}

\bsp	
\label{lastpage}
\end{document}